%% file: fieldNox.tex
\documentclass[review]{elsarticle}

\usepackage[utf8]{inputenc}
\usepackage[T1]{fontenc}
\usepackage{tgtermes}
\usepackage{lineno,hyperref,pifont,natbib,geometry,fleqn,graphicx,makeidx,fancyhdr,multirow,verbatim,amsmath,amsthm,amssymb,float,array,subfloat,epstopdf,subfig,upgreek,appendix,placeins}
\modulolinenumbers[5]

\journal{Journal of Instrumentation}

\begin{document}

\begin{frontmatter}

\title{A method to observe field-region oxide charge and inter-electrode isolation from $CV$-characteristics of $n$-on-$p$ devices}


\author[a]{T. Abdilov}
\author[a]{N. Akchurin}
\author[a]{C. Carty}
\author[a]{Y. Kazhykarim}
\author[b]{V. Kuryatkov}
\author[a]{T. Peltola\corref{cor2}}
\ead{timo.peltola@ttu.edu}
\author[h]{A. Wade}

\address[a]{Department of Physics and Astronomy, Texas Tech University, 1200 Memorial Circle, Lubbock, Texas, U.S.A.}
\address[b]{Nanotech Center, Texas Tech University, 902 Boston Ave, Lubbock, Texas, U.S.A.}
\address[h]{Department of Physics, Florida State University, 77 Chieftan Way, Tallahassee, Florida, U.S.A.}
\cortext[cor2]{Corresponding author}




\begin{abstract}
\label{Abstract}
\input{Abstract.tex}

\end{abstract}

%
\end{frontmatter}


\newpage
\section{Introduction}
\label{Introduction}
\input{Introduction.tex}
\section{Test-structure samples and HGCAL sensors}
\label{TdS}
\input{SamplesFluences.tex}
\section{Measurement and simulation setups}
\label{Msetup}
%
\input{setups.tex}
%

%
\section{Experimental and simulation results}
\label{Results}
%
\input{Results2.tex}
\section{Discussion on radiation damage} 
\label{Discussion}
\input{discussion_fT.tex}

\section{Summary and conclusions}
\input{Summary.tex}
\label{Summary}

\section*{Acknowledgements}
This work has been supported by the US Department of Energy, Office of Science (DE-SC0015592). 
We thank Ronald Lipton of FNAL for his invaluable insights on the analysis of this investigation. 
We also thank K. Zinsmeyer, C. Perez, and posthumously P. Cruzan of TTU for their expert technical support. 
%

\bibliography{mybibfile}

\newpage
\appendix
\label{Appendix}
\input{appendices.tex}

\end{document}

%% file: Abstract.tex
$N$-on-$p$ silicon sensors will be 
utilized in the 
Compact Muon Solenoid (CMS) detector's 
tracker and High Granularity Calorimeter (HGCAL) in the 
High Luminosity upgrade of the Large Hadron Collider (HL-LHC). Among their several advantages in terms of radiation hardness over the traditional $p$-on-$n$ sensors in the extreme radiation environment of the HL-LHC 
are electron collection instead of holes and overlapping maxima of weighting and electric fields at the 
charge-collecting electrodes. 
%
%
The disadvantage of the multi-channel SiO$_2$-passivated $n$-on-$p$ sensors is the generation of an inversion layer under the Si/SiO$_2$-interface by a 
positive interface-oxide-charge ($N_\textrm{ox}$) that at high densities can compromise the position resolution by creating a conduction channel between the electrodes. This issue is typically addressed by including additional isolation implants ($p$-stop, $p$-spray) between $n^+$-electrodes.
%
Focusing on the guard-ring regions of $n$-on-$p$ sensors 
where no isolation implants are applied between the electrodes, a capacitance-voltage ($CV$) characterization study of both 6-inch wafer test diodes and 8-inch HGCAL prototype and pre-series sensors showed a distinct threshold voltage ($V_\textrm{th,iso}$) in the $CV$-characteristics of a biased $n^+$-electrode when its enclosing guard-ring was left floating. When reproduced by simulations, the measured $V_\textrm{th,iso}$ was found to contain information on 
the field-region 
$N_\textrm{ox}$ 
and indicate the threshold where the two electrodes become electrically isolated by the influence of the reverse bias voltage.
Together with previous studies on the inter-electrode isolation of irradiated $n$-on-$p$ sensors, the 
results 
indicate 
that position sensitive $n$-on-$p$ sensors without isolation implants may be feasible in future HEP experiments. 

%% file: Introduction.tex
In order to cope with the extreme radiation environments 
of the Large Hadron Collider and future hadron colliders like FCC-hh\footnote{Future Circular Hadron Collider} and SPPC\footnote{Super Proton-Proton Collider}, 
one extensively studied approach to improve the radiation hardness of planar silicon particle detectors is the application of $n^{+}/p^{-}/p^+$ ($n$-on-$p$, $n$-in-$p$) structures
instead of the conventional $p$-on-$n$ configuration \cite{Casse2002p,Harkonen2007b}. Since radiation introduces acceptor-like bulk defects, 
no type inversion occurs in the $p$-type bulk resulting in a favorable combination of the maxima of the weighting and electric fields after
irradiation, which maximizes the signal and charge carrier velocity in the vicinity of the collecting electrode \cite{Hartmann2017}. The readout at $n^+$-electrodes 
results in a signal 
generated by the drift of electrons, and since electrons have about three times higher mobility and longer trapping
times than holes, the amount of trapped charge carriers during their drift is reduced. This enables a high-speed readout and higher 
collected charge in $n$-on-$p$ devices than in conventional sensors. Also, after irradiation the electric fields at the $n^+$-electrodes are lower than in $p$-on-$n$ sensors, resulting in 
better noise performance \cite{Casse2002p,Adam2017}. 
Another asset of the $n$-on-$p$ sensor is the reduced dependence of the charge collection on the reverse annealing of the effective space charge in highly irradiated detectors \cite{Casse2006p,Kramberger2002p}.
%
As a consequence, $n$-on-$p$ sensors 
are the baseline technology for the 
detector upgrades in, for example, the Compact Muon Solenoid (CMS) experiment's 
tracker and 
High Granularity Calorimeter (HGCAL) \cite{Phase2}, where the replacement of the 
endcap calorimeter will involve about 600 m$^2$ of 8-inch silicon modules \cite{Brondolin2020,TQ2023} that will need to sustain fluences of about $1\times10^{16}~\textrm{n}_\textrm{eq}\textrm{cm}^{-2}$\footnote{1-MeV neutron equivalent fluence} and doses in excess of 1.5 MGy.

The disadvantage of $n$-on-$p$ devices is the more complex fabrication technology due to the required isolation structures between the $n^+$-electrodes. For example, conventional AC-coupled, poly-Si biased, $p$-on-$n$ sensors can be processed with 6--7 lithography mask levels, while $n$-on-$p$ devices require two more lithography steps and additional ion implantations \cite{Harkonen2007b}. The isolation implants are utilized to counter the effect from the inherent positive charge that is introduced by impurities during the growth process of SiO$_2$ inside the oxide and at the Si/SiO$_2$-interface ($N_\textrm{ox}$). The positive $N_\textrm{ox}$ attracts electrons (minority carriers in $p$-type bulk) from the Si-bulk that form an inversion layer under the interface which 
acts as a conduction channel between the segmented $n^+$-electrodes. 
Radiation induced accumulation of surface damage can lead to a substantial increase of net $N_\textrm{ox}$ \cite{Peltola2023}. The negative space charge of the additional highly $p$-doped 
implant between the $n^+$-electrodes breaks the 
inversion layer by suppressing the electron density 
at the vicinity of Si/SiO$_2$-interface, thus preserving the inter-electrode isolation and position resolution. Two solutions for the isolation implants are 
commonly utilized; $p$-stop and $p$-spray (or the combination of the two, `moderated' $p$-spray \cite{Hartmann2017}), with localized and uniformly distributed doping concentrations, respectively, along the surface in the inter-electrode gap \cite{Kemmer1993,Richter1996,Iwata1998,Verzellesi2000,Piemonte2006,Unno2013,Printz2016}. For stable device performance, careful tuning of the processing parameters is needed, since high doping concentrations in the isolation implants ($\gtrsim5\times10^{16}~\textrm{cm}^{-3}$) increase the probability of discharges or avalanche effects due to excessive localized electric fields \cite{Adam2017,Printz2016,Dalal2014b}. 
For the same reason the actual contact between the isolation and $n^+$ implants should also be avoided \cite{Hartmann2017}.

However, previous studies have reported high levels of inter-electrode isolation in $n$-on-$p$ microstrip sensors without isolation implants before and after irradiations with protons \cite{Unno2007,Gosewich2021_J} and neutrons \cite{Gosewich2021_J}. Thus, to investigate further possibilities of planar $n$-on-$p$ sensor configurations, this study focuses on the properties of the regions of $n$-on-$p$ devices without isolation implants between $n^+$-electrodes, i.e., the region between a diode and its guard-ring, and the observed threshold voltage of inter-electrode isolation ($V_\textrm{th,iso}$). $V_\textrm{th,iso}$ introduced in this paper is the threshold voltage where the applied reverse bias voltage becomes sufficiently high to sweep electrons from the inter-electrode gap to the positively biased n$^+$-electrode to the extent that breaks the conduction channel and electrically isolates the electrodes. Simultaneous to this, $V_\textrm{th,iso}$ provides information on the level of $N_\textrm{ox}$ in the field-region of an $n$-on-$p$ sensor. 

Two substantially different device types were included in the study: test diodes from a 6-inch HGCAL prototype wafer and the guard-ring periphery region of hexagonal 8-inch HGCAL multi-channel sensors.
An accompanying study \cite{Peltola2023} 
discusses 
the mechanisms behind the 
beneficial impact on 
inter-electrode isolation introduced by irradiation. 

The paper is arranged 
by first introducing the test-structure samples from 6-inch wafers and the 8-inch HGCAL-prototype and pre-series sensors 
in Section~\ref{TdS}. Next, the measurement and simulation setups are described in Section~\ref{Msetup}. 
The experimental and simulated $CV$-characterization results start in Section~\ref{TDs} with the observations and interpretation of 
$V_\textrm{th,iso}$ in the test-diode $C^{-2}V$-curves, and the investigation of its relation to the oxide charge density at the Si/SiO$_2$-interface ($N_\textrm{ox}$), active thickness of the sensor ($t$) and Si-bulk doping ($N_\textrm{B}$), as well as the correlation of $V_\textrm{th,iso}$ to the inter-electrode isolation of $n^+$-electrodes without isolation implants. 
The results of 8-inch HGCAL sensors are presented in Section~\ref{HGCALsensors} by first determining the field-region $N_\textrm{ox}$ for the various sensor $t$ and configurations, 
followed by the investigation 
of $V_\textrm{th,iso}$ of inter-electrode isolation with $N_\textrm{ox}$.  
Finally, the results and radiation damage effects are discussed in Section~\ref{Discussion}, while summary and conclusions 
are given in Section~\ref{Summary}. 

%% file: SamplesFluences.tex
\begin{figure}[htb!]
\centering
\subfloat[]{\includegraphics[width=.5\textwidth]{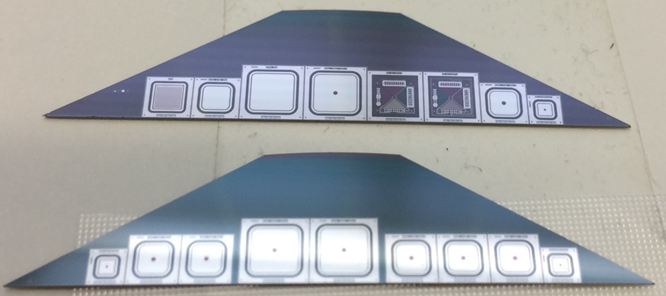}\label{TD_MOS_6in}}\hspace{1mm}%
\subfloat[]{\includegraphics[width=.38\textwidth]{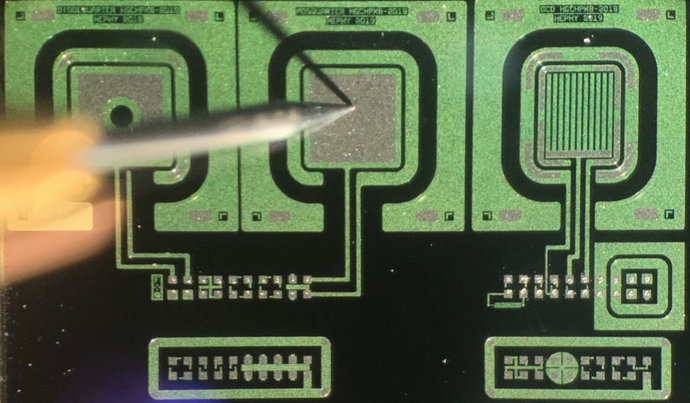}\label{MOS_8in}}
\caption{\small 
Test structure samples. (a) `Half-moon' samples with test structures diced from a 6-inch wafer \cite{Peltola2020}. Top: three test diodes,
two MOS-capacitors (second and third from left) and other test structures. Bottom: nine test diodes. (b) Close-up of a test-structure sample diced from an 8-inch HGCAL wafer with the MOS-capacitor connected to the measurement circuit by a probe needle. A test diode enclosed by its guard-ring is visible on the left-side of the MOS-capacitor. 
}
\label{Samples}
\end{figure}

The three `half-moon' samples in the study with `Large', `Half' and `Quarter' 
test-diodes and MOS\footnote{Metal-Oxide-Semiconductor}-capacitors (with active areas of $4.0\times4.0$, $2.5\times2.5$ and $1.25\times1.25~\textrm{mm}^2$, respectively) 
were diced from a 6-inch Hamamatsu Photonics K.K. (HPK) 
320-$\upmu\textrm{m}$-thick $n$-on-$p$ sensor-wafer\footnote{Early prototype wafer for the HGCAL.}. 
The test structures on the samples in Figure~\ref{TD_MOS_6in} were designed at Institut f{\"u}r Hochenergiephysik (HEPHY). 
The standard-diffusion process was utilized in the 
heavily doped backplane blocking contact 
to produce 300-$\upmu\textrm{m}$ active thickness test diodes. 
More extensive details of the test-diode design and bulk properties are given in \cite{Peltola2020}. 

\begin{figure}[htb!]
\centering
\subfloat[]{\includegraphics[width=.4\textwidth]{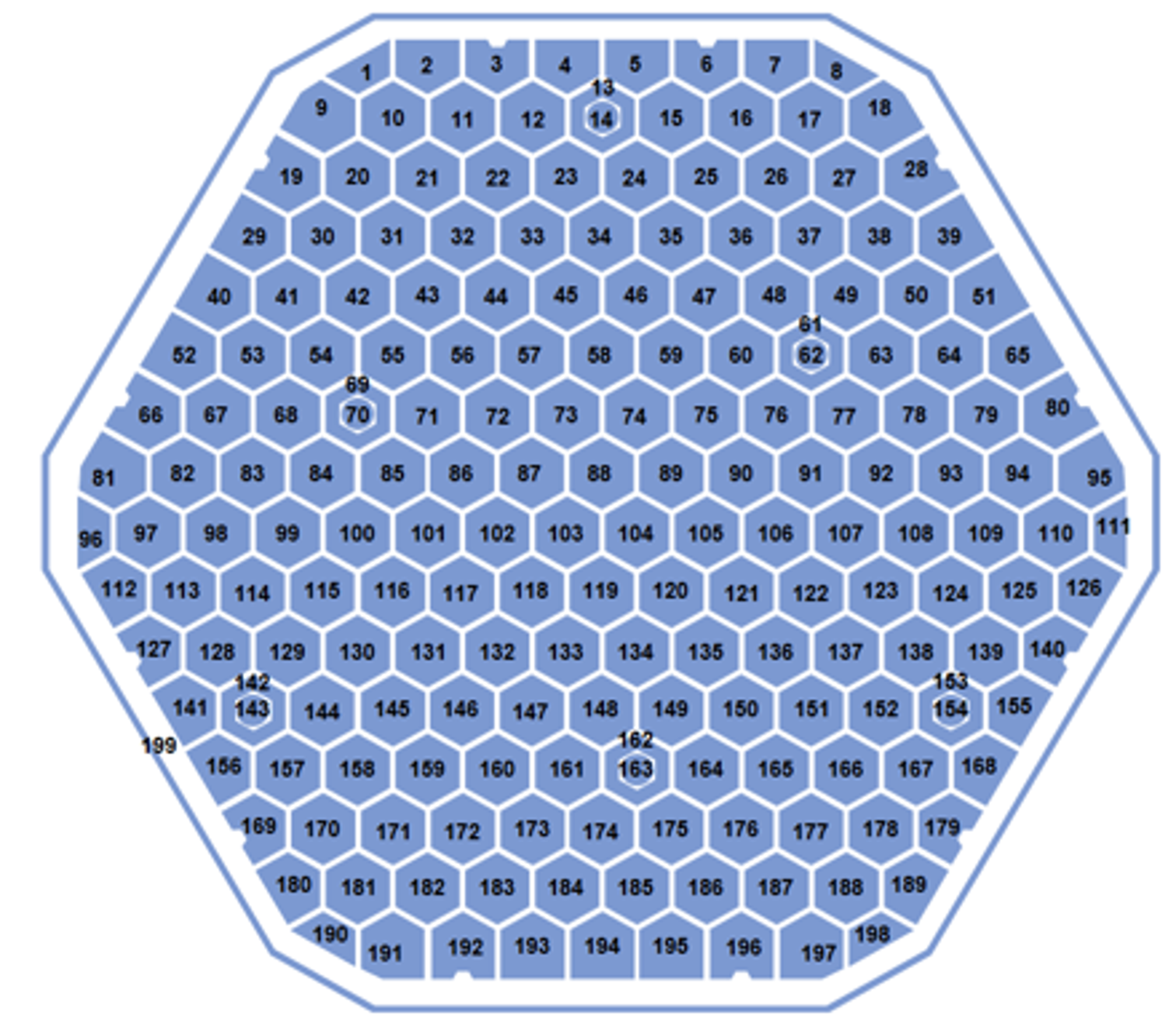}\label{LDfull}}\hspace{1mm}%
\subfloat[]{\includegraphics[width=.4\textwidth]{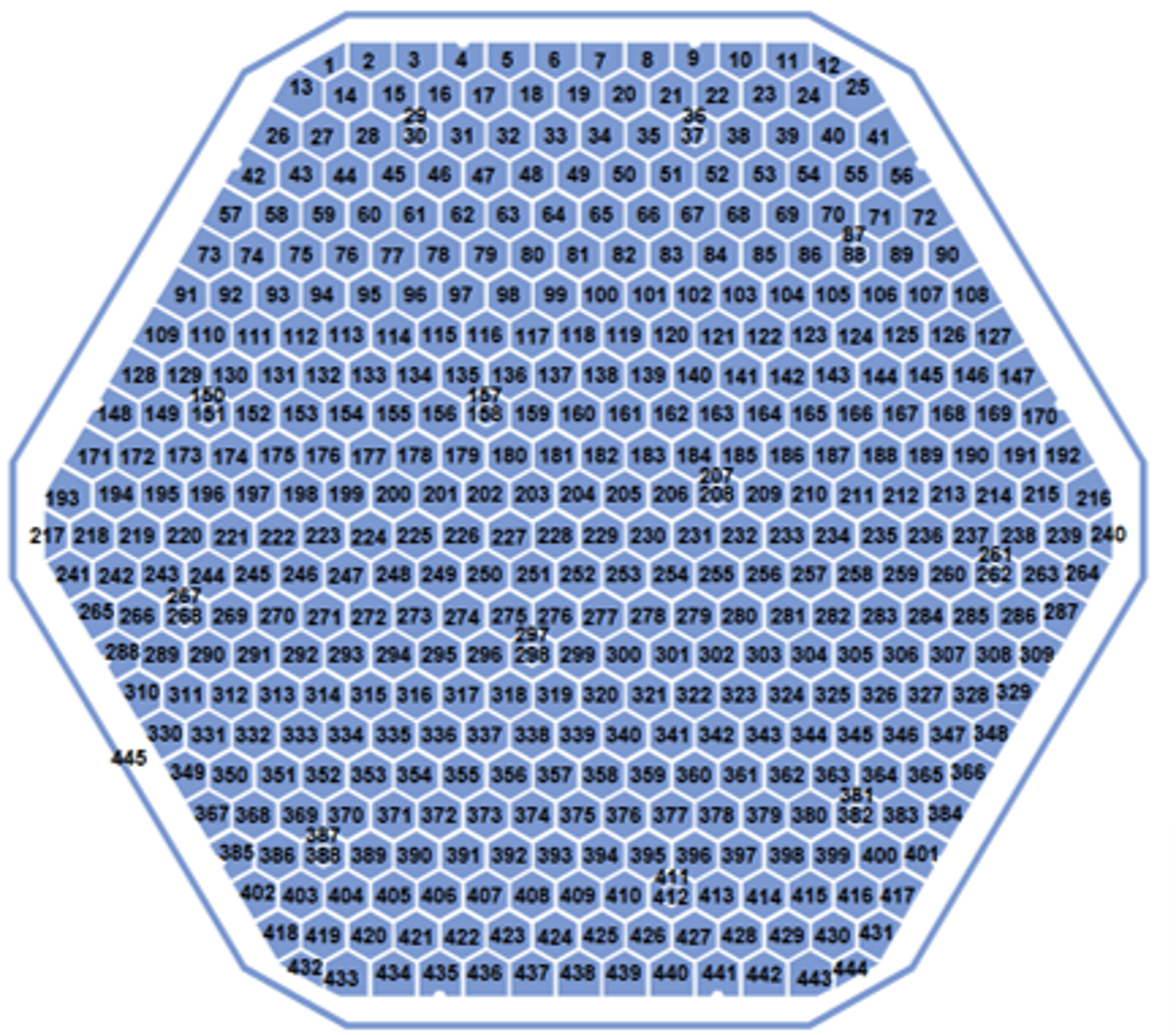}\label{HDfull}}\\%
\subfloat[]{\includegraphics[width=.4\textwidth]{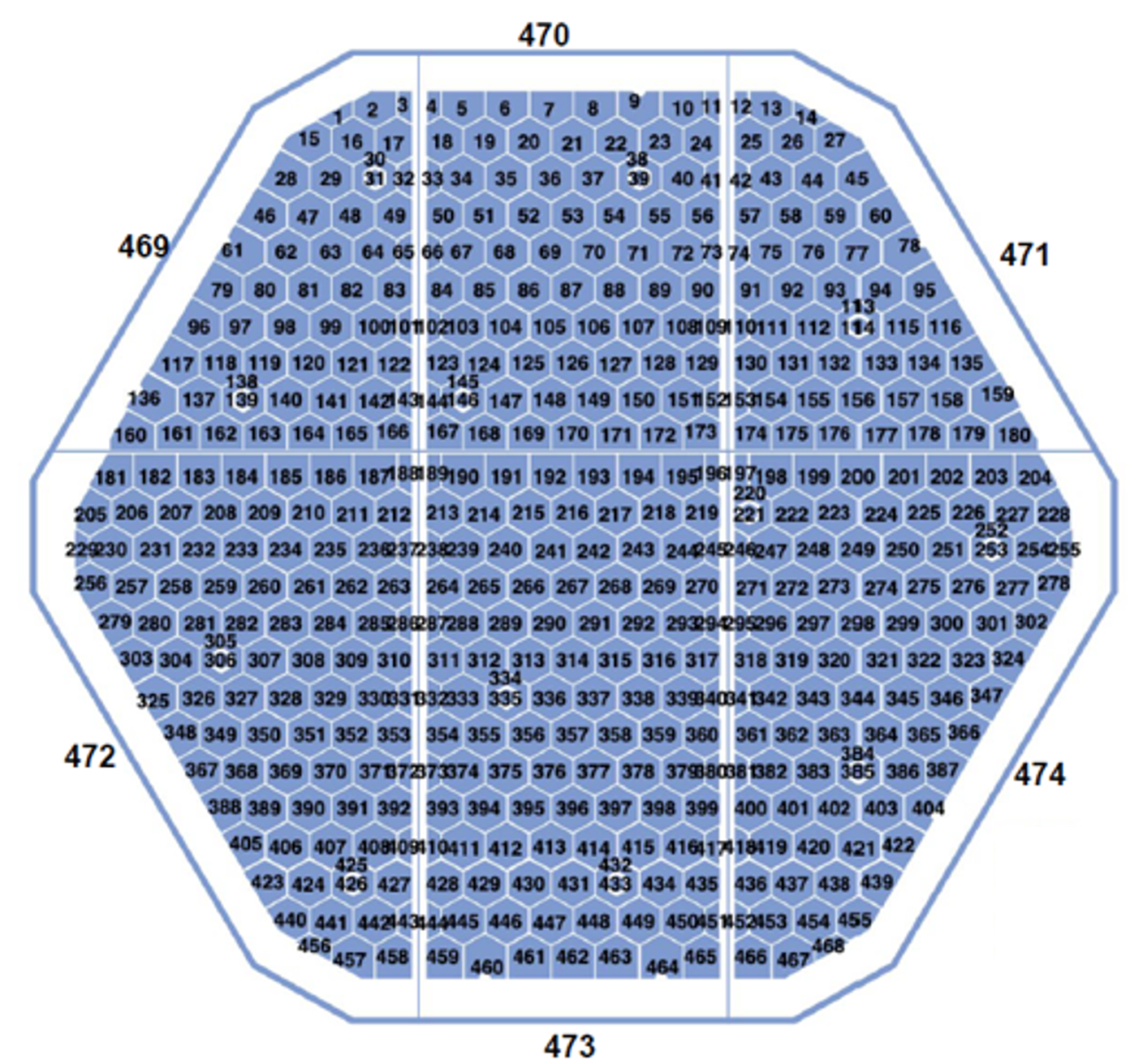}\label{HDmgw}}%
\caption{\small HGCAL 8-inch multi-channel sensor designs included in the study. (a) Full low-density (LD) float-zone Si-sensor with 198 
channels (pre-series). (b) Full high-density (HD) epitaxial Si-sensor with 444 
channels (pre-series). The cell area of the HD sensors is less than half of the area of the LD-sensor cells to control the cell capacitance. (c) Multi-Geometry Wafer (MGW) HD epitaxial Si-sensor (version 2 prototype). Boundary of the `bottom-cut' partial sensor with 288 
channels is shown by the horizontal blue line in the middle part of the sensor. The channels of the full sensors are enclosed by a single biased guard-ring (channels 199 and 445 in Figures~\ref{LDfull} and~\ref{HDfull}, respectively), while the channels of the `bottom-cut' partial HD-sensor are divided into three regions by three biased guard-rings (channels 472, 473 and 474 in Figure~\ref{HDmgw}).
}
\label{Sensors}
\end{figure}

The HGCAL version 2 prototype (or `proto-A') and pre-series\footnote{Consecutive advanced HPK production stages towards the final HGCAL sensor production parameters.} 8-inch multi-channel $n$-on-$p$ sensors\footnote{For shortness, the prototypes will be referred to as `HGCAL sensors' in the following sections.}, designed at Fermi National Accelerator Laboratory (FNAL) and produced by HPK, included in the study are presented in Figure~\ref{Sensors}. 
The low-density (LD) sensors included both 300- and 200-$\upmu\textrm{m}$ thicknesses, which were produced from physically thinned $p$-type float zone silicon wafers, resulting in essentially equal active and physical thicknesses. The high-density (HD) sensors with an active thickness of 120 $\upmu\textrm{m}$ were produced by growing epitaxial
silicon on a lower resistivity substrate, resulting in a total physical thickness of 300 $\upmu\textrm{m}$ \cite{Phase2}. The $n^+$-electrodes of the several hundred sensor cells (= channels) displayed in Figure~\ref{Sensors} are isolated from each other by `common' $p$-stop implants, i.e., one $p$-stop divides the inter-cell gap (as opposed to `individual' $p$-stop where each cell has a dedicated $p$-stop). Corresponding `Quarter' MOS-capacitors from the 8-inch sensor wafers in Figure~\ref{MOS_8in} designed at HEPHY 
were also included in the electrical characterization investigation. 

\begin{figure}[htb!]
\centering
\includegraphics[width=1.0\textwidth]{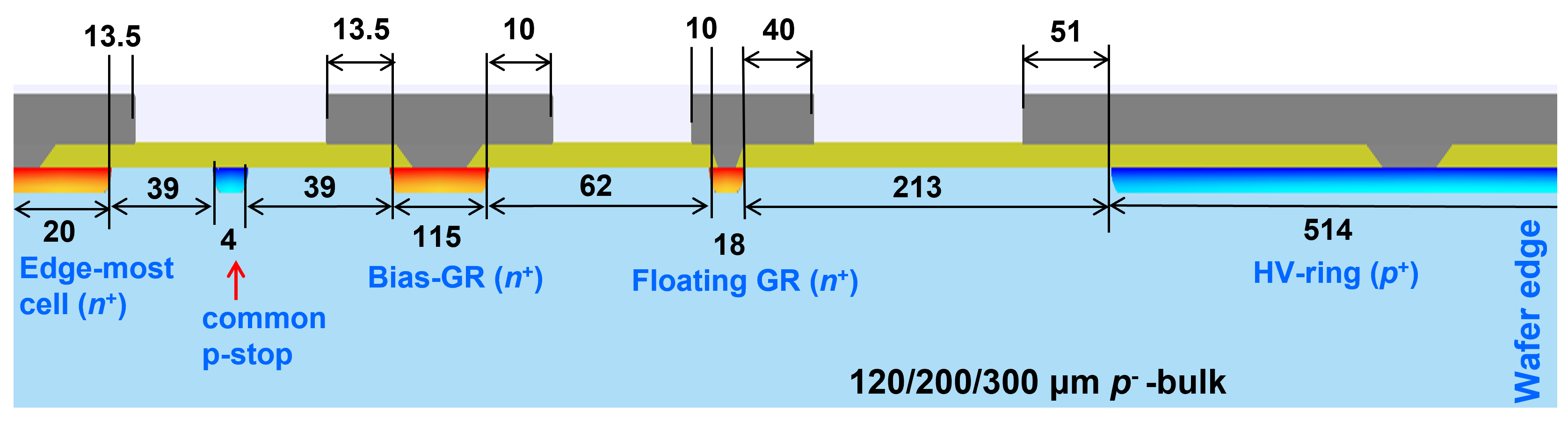}
\caption{\small Cross-sectional sketch of the front-side of the periphery region of an HGCAL sensor. The horizontal dimensions are in $\upmu\textrm{m}$. Gray: 1-$\upmu\textrm{m}$-thick Al-layer for DC-coupled electrical contacts, dark yellow: 700-nm-thick SiO$_2$ layer. The backplane of the sensor has the $p^{+}$-blocking contact with 1-$\upmu\textrm{m}$-thick uniform Al-layer. The biased guard-ring (bias-GR) is grounded during the measurements of the cells of the multi-channel sensor, 
while the high voltage (HV)-ring is used for the front-side biasing scheme \cite{Baselga2018}, where the applied $-$HV connects to the backplane via the low-resistivity edge-region under the HV-ring (with a $p^{+}$/$p^{-}$/$p^{+}$ configuration). The $CV$-characterizations in this study were done on the bias-GR.}
\label{HGCALedge}
\end{figure}

%
Both the test structures and the HGCAL sensors have $\langle100\rangle$ crystal orientation, and 
about 700-nm-thick silicon dioxide (SiO$_2$) as the passivation layer
with inherent positive oxide charge. The gap between the $n^{+}$-implant of the test diode and its guard-ring implant is 100 $\upmu\textrm{m}$ for all three test-diode sizes, while the periphery region of the HGCAL sensors is presented in Figure~\ref{HGCALedge}.  
%
%
%
%
%

%% file: setups.tex
%
For the 
electrical characterizations of both the test structures and the multi-channel HGCAL sensors, two measurement setups were utilized at Texas Tech University (TTU).
The capacitance-voltage ($CV$) characterizations 
of the test structures were carried out with a single-channel 
custom probe station. 
The reverse bias voltage of the test diode (or the gate voltage of the MOS-capacitor) was supplied by a Keithley 2410 source measure unit (SMU) 
while the 
capacitance was read out by a 
Keysight E4980AL LCR-meter. 
The 
negative high voltage ($-$HV) was provided to the backplane of the test structure by a vacuum chuck, 
and the connection to the measurement circuit was 
realized with a probe needle on the segmented front surface, as shown in Figure~\ref{MOS_8in}. A second probe needle was applied for the grounding of the test-diode's guard-ring.
The setup is described in detail in refs. \cite{Peltola2023,Peltola2020}. 

The corresponding $CV$-characterizations of the HGCAL sensors were carried out with a multi-channel setup utilizing an automated ARRAY dual-card setup \cite{Brondolin2019} that enables the electrical characterizations of the 8-inch sensors containing several hundred individual cells. For the connection to the measurement circuit a dedicated probe card for each sensor type in Figure~\ref{Sensors} was required. 
Probe cards for full sensors in Figures~\ref{LDfull} and~\ref{HDfull} and for HD-partial sensor in Figure~\ref{HDmgw} utilized front-side and back-side biasing, respectively. The guard-ring next to the HV-ring in Figure~\ref{HGCALedge} is left floating in all probe-card types. The remote control and 
data acquisition functions 
were realized with a LabVIEW\texttrademark-based custom program (HexDAQ version 1.7\footnote{https://gitlab.cern.ch/CLICdp/HGCAL/}). The same SMU and LCR-meter units were used for both single-channel and multi-channel circuit measurements. 
 
The 2D device-simulations 
were carried out using the Synopsys Sentaurus\footnote{http://www.synopsys.com} finite-element Technology Computer-Aided Design (TCAD) software framework. 
For the test-diode and MOS-capacitor simulations, a structure consisting of a DC-coupled $n^+$-pad between two $n^+$ guard-rings (GR) (either grounded or floating), and a metal-oxide-silicon structure, respectively, were implemented. The backplane doping profile and oxide thickness 
were set to match the real test structures as described in Section~\ref{TdS} and in \cite{Peltola2020}. 

\begin{figure*}
\centering
\includegraphics[width=.7\textwidth]{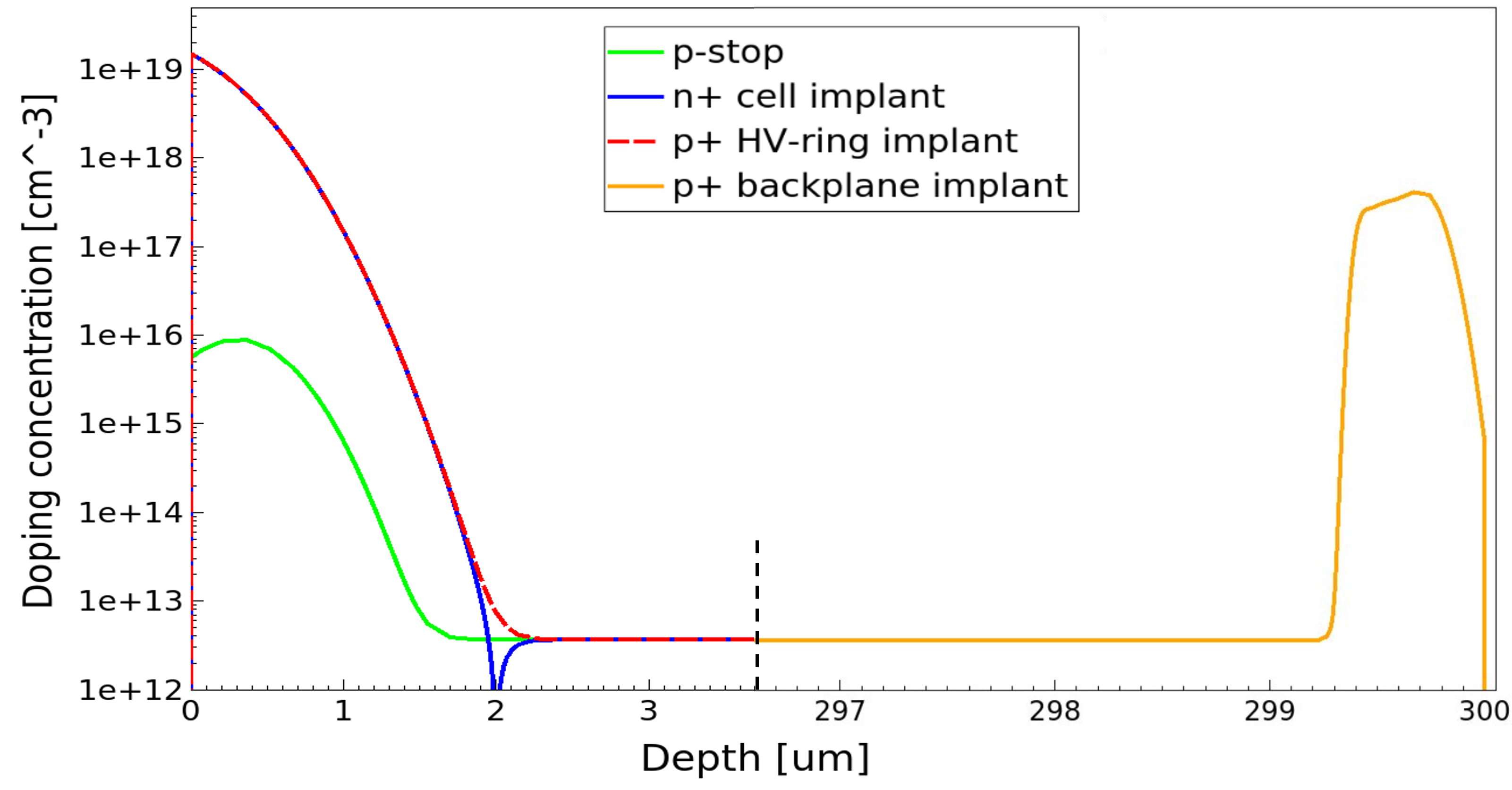}
\caption{\small Input doping profiles for the 300-$\upmu\textrm{m}$-thick HGCAL LD-sensor periphery-region TCAD-simulation. Identical front-side ($\textrm{Depth}=0~\upmu\textrm{m}$) doping profiles were set for the HD-sensors, while the low-resistivity substrate of the epitaxial sensor was modeled by an error function starting from the backplane ($\textrm{Depth}=300~\upmu\textrm{m}$) with an abrupt drop to bulk-doping level at 120-$\upmu\textrm{m}$ depth.}
\label{HGCALdoping}
\end{figure*}

For the HGCAL sensor simulations of the three active thicknesses, the periphery-region with the dimensions displayed in Figure~\ref{HGCALedge} was implemented. 
Results from spreading resistance profiling (SRP) measurements of the sensors carried out within the HGCAL community were used as an input for the doping profiles, shown in Figure~\ref{HGCALdoping}. The bulk-doping ($N_\textrm{B}$) levels were extracted from the $CV$-measured full-depletion voltages ($V_\textrm{fd}$) of the sensor-cells shown in~\ref{App1} 
and implemented in the simulation. 
%
%

%% file: Results2.tex
\subsection{6-inch wafer test structures}
\label{TDs}
The room-temperature $CV$-characterizations of the test structures were carried out at a frequency of 200 kHz to 
control the dissipation factor and stabilize 
the measurements (reduced capacitive impedance \cite{Stauffer2008}) of the low-capacitance test diodes with a minimum capacitance of 0.54 pF (`Quarter' diode). 

\begin{figure}[htb!]
     \centering
     \subfloat[]{\includegraphics[width=.49\textwidth]{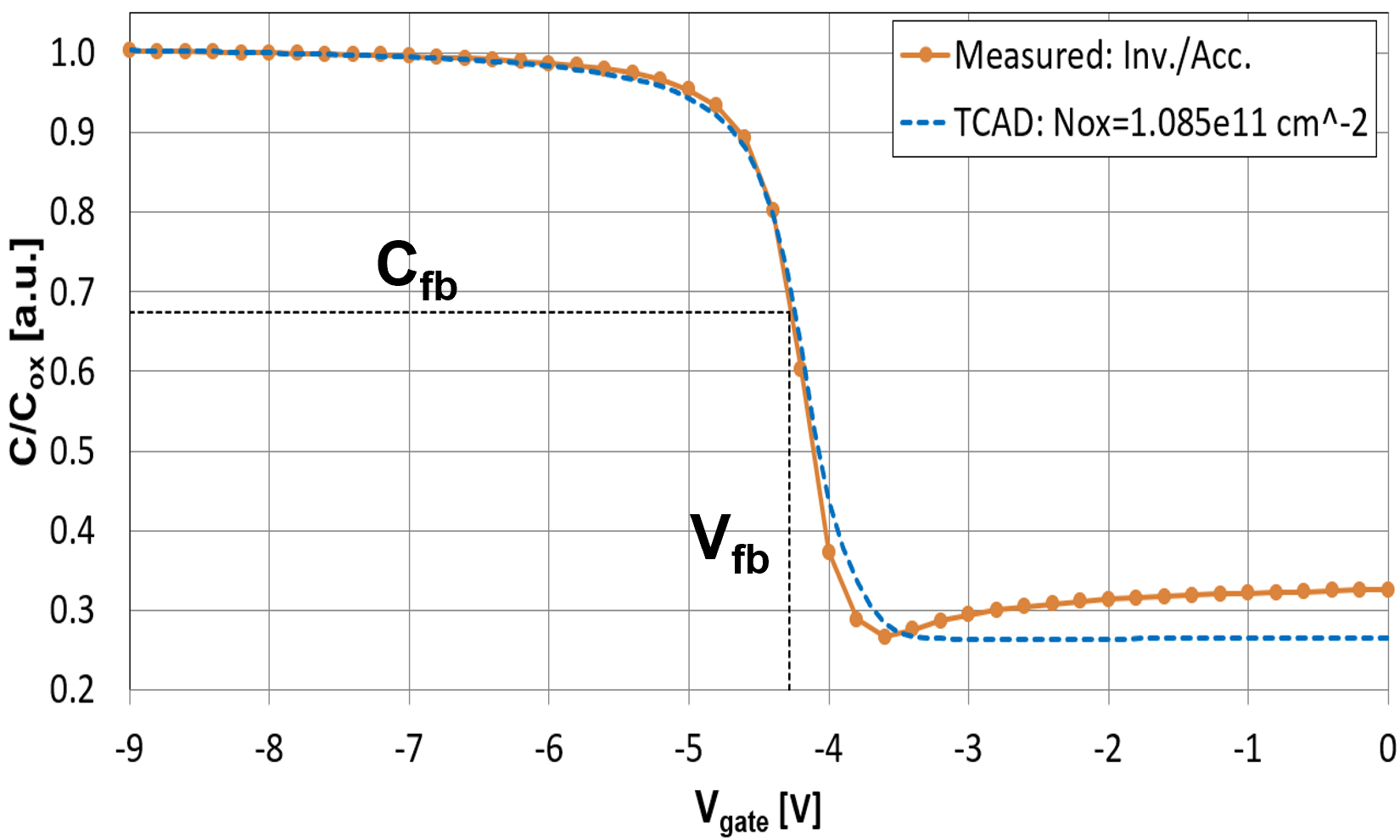}\label{MOS}}\hspace{1mm}%
     \subfloat[]{\includegraphics[width=.495\textwidth]{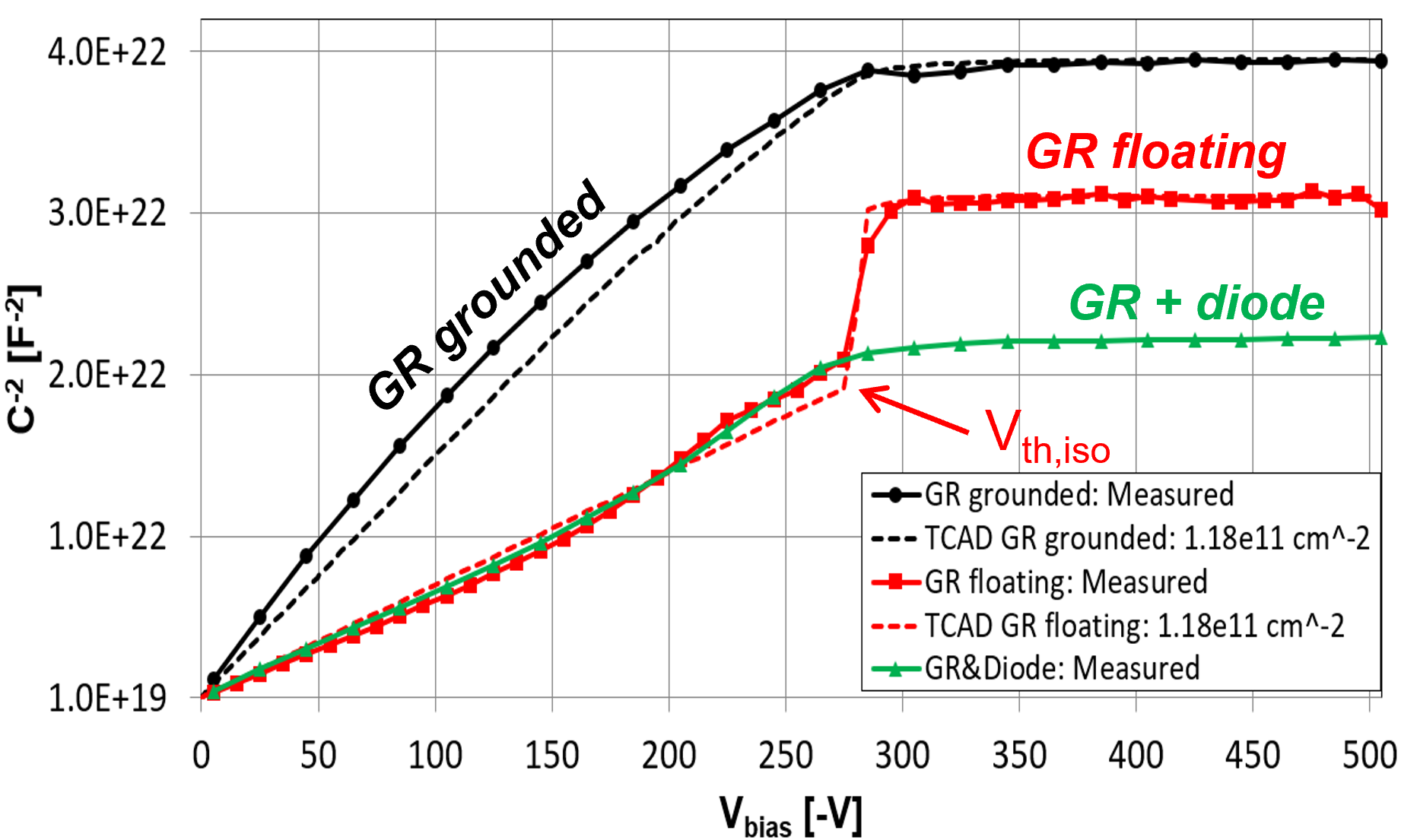}\label{TD_gnd_float}}\\
     \subfloat[]{\includegraphics[width=.495\textwidth]{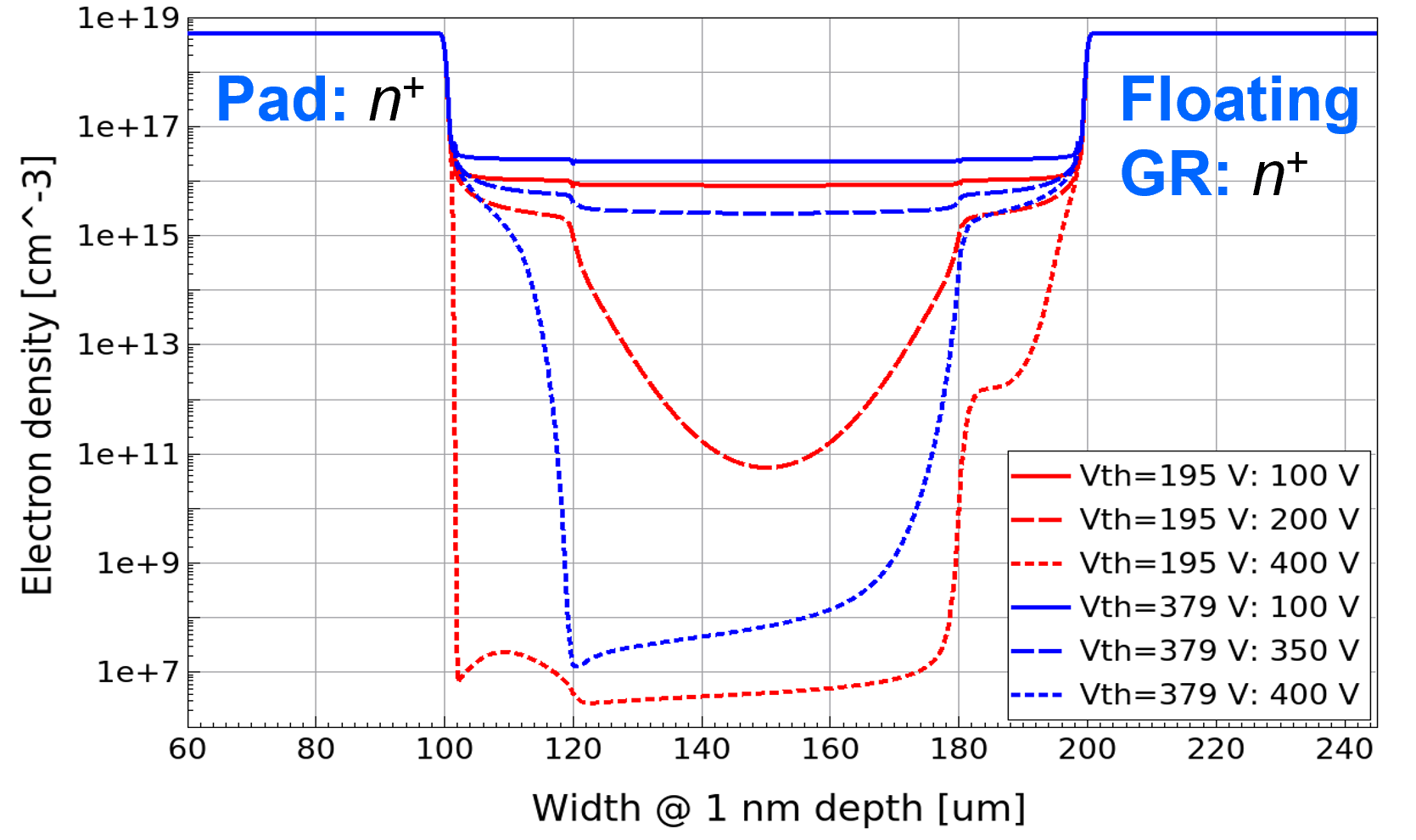}\label{eD_PadToGR}}
    \caption{\small (a) Measured and TCAD-simulated $CV$-characteristics of a $p$-bulk MOS-capacitor. 
    Measured $CV$-curves were identical for the $CV$-sweeps starting either from inversion (Inv.) or accumulation (Acc.) regions. 
    (b) Measured and simulated $CV$-characteristics of an $n$-on-$p$ test diode, with its guard-ring (GR) either grounded, floating or 
    measured together with the test diode (GR + diode). The point of 
    $V_\textrm{th,iso}$ reproduced by the simulation with $N_\textrm{ox}=1.18\times10^{11}~\textrm{cm}^{-2}$ is indicated. 
    The measured test structures in 
    (a) and (b) were from the same 6-inch Si-wafer. (c) Simulated evolution of electron density with reverse bias voltage at 1-nm depth from the Si/SiO$_2$-interface and between the test-diode pad and its floating guard-ring. The two values of $V_\textrm{th,iso}$, 195 and 379 V, correspond to $N_\textrm{ox}$ values of $1.0\times10^{11}$ and $1.4\times10^{11}~\textrm{cm}^{-2}$ in Figure~\ref{Vth_measTCAD}, respectively. The metal-overhang edges of the two electrodes are at 120 and 180 $\upmu\textrm{m}$, respectively. 
}
\label{MOS_TD}
\end{figure}

The measured $CV$-characteristics of the `Large' MOS-capacitor and test diode are presented in Figure~\ref{MOS_TD}. 
The flat band voltage ($V_\textrm{fb}$) of the MOS-capacitor was determined by the flat-band-capacitance method ($C_\textrm{fb}$ in Figure~\ref{MOS}). 
The extracted $V_\textrm{fb}=-4.28\pm0.02~\textrm{V}$ was then used to resolve the oxide charge density at the Si/SiO$_2$-interface as $N_\textrm{ox}=(1.08\pm0.05)\times10^{11}~\textrm{cm}^{-2}$ 
\cite{Peltola2023}. It was also shown in \cite{Peltola2023} that in the absence of hysteresis whether the $CV$-sweep of the pre-irradiated MOS-capacitor is initiated from deep accumulation or from deep inversion ($-9~\textrm{V}$ and $0~\textrm{V}$ in Figure~\ref{MOS}, respectively), the contribution from mobile ionic charges ($N_\textrm{M}$) and interface traps ($N_\textrm{it}$) to $N_\textrm{ox}$ is negligible, leading to $N_\textrm{ox}$ being described essentially by the fixed oxide charge density ($N_\textrm{f}$), i.e., $N_\textrm{ox}~{\cong}~N_\textrm{f}$. As shown in Figure~\ref{MOS}, the measured $V_\textrm{fb}$ was reproduced by the simulation with the input value of $N_\textrm{ox}=N_\textrm{f}$ (implemented at the Si/SiO$_2$-interface as a positive charge-sheet 
with a uniform distribution along the interface) within uncertainty of the experimentally extracted $N_\textrm{ox}$ (with contributions to uncertainty from measured $V_\textrm{fb}$ and oxide capacitance ($C_\textrm{ox}$)). Therefore, this approach is maintained for the field-region $N_\textrm{ox}$ 
simulations 
on the test diodes and HGCAL sensors (in the following sections the simulation tuning-parameter at the Si/SiO$_2$-interface will be referred to as $N_\textrm{ox}$). 

Shown in Figure~\ref{TD_gnd_float}, the 
characteristic $C_\textrm{geom}$-values --where capacitance is dependent only on the material and the geometry of the device-- and the slopes in the dynamic-region of the measured $C^{-2}V$-curves 
from either solely the test-diode pad (`GR grounded') or both the pad and its guard-ring (`GR + diode') reflect the differences in the active areas $A$ involved in the measurements by the relation
\begin{linenomath}
\begin{equation}\label{eq1}
C_\textrm{geom}=\epsilon_\textrm{s}\frac{A}{d},
\end{equation}
where $\epsilon_\textrm{s}$ is the product of vacuum permittivity and the relative permittivity of silicon, and 
$d$ is the active thickness of the device 
\cite{Sze1981}. Hence, the measured capacitances are $C=C_\textrm{diode}$ and $C'=C_\textrm{diode}+C_\textrm{GR}=C_\textrm{tot}$ for the two configurations throughout the applied voltage range.
\end{linenomath}

However, the measurement with the floating guard-ring (i.e., guard-ring is disconnected from the biasing circuit, thus its potential is 
`floating' with changing conditions in the device) in Figure~\ref{TD_gnd_float} displays a double slope in the dynamic-region of $C^{-2}V$-curve with a distinct threshold voltage ($V_\textrm{th,iso}$) between the two slopes. The identical slope with $C_\textrm{tot}$-curve at $V<V_\textrm{th,iso}$ suggests that the diode and its guard-ring are shorted in this voltage-region. This is caused by the conduction channel between the test-diode pad and the guard-ring --where there is no isolation implant--, formed by the positive oxide charge ($Q_\textrm{ox}$) of SiO$_2$ attracting minority carriers, i.e., electrons, from Si-bulk to the Si/SiO$_2$-interface. As shown by the simulated evolution of the electron density in the vicinity of the interface in Figure~\ref{eD_PadToGR}, at $V~{\geq}~V_\textrm{th,iso}$ the Coulomb force from the positively biased $n^+$-pad of the test diode 
overcomes the attraction from $Q_\textrm{ox}$, and the electrons are swept from the inter-electrode gap (electron density drops by about nine-orders-of-magnitude from $V<V_\textrm{th,iso}$) to the $n^+$-pad, 
resulting in the test diode becoming isolated from the guard-ring. Due to this process, the area involved in the $CV$-measurement of the test-diode pad is reduced, which, as a consequence of Eq.~\ref{eq1}, is observed as an abrupt decrease of $C$ in Figure~\ref{TD_gnd_float}, forming a double-slope in the dynamic-region of the $CV$-curve.

\begin{figure}[htb!]
     \centering
     \subfloat[]{\includegraphics[width=.495\textwidth]{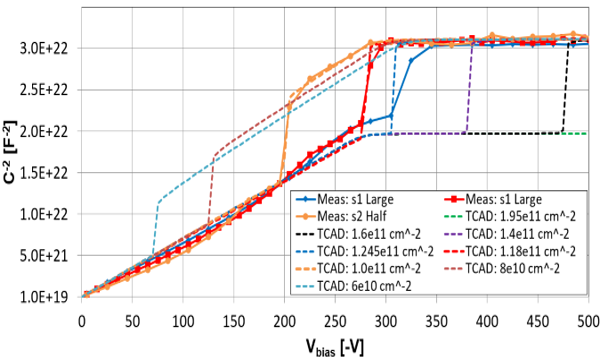}\label{Vth_measTCAD}}\hspace{1mm}%
     \subfloat[]{\includegraphics[width=.495\textwidth]{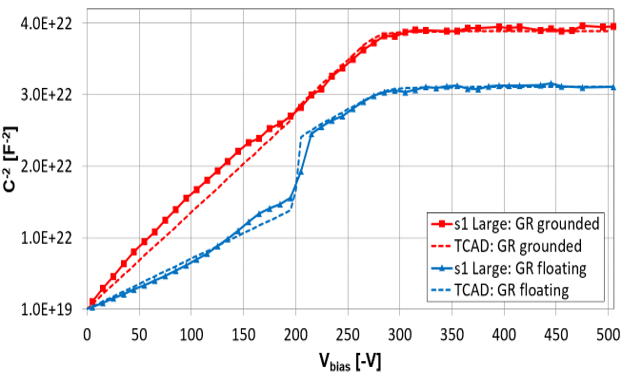}\label{s1L_gnd_float}}
    \caption{\small Measured and simulated $CV$-characteristics of 300-$\upmu\textrm{m}$-thick $n$-on-$p$ test diodes. `s1'=`half-moon' sample one. (a) Influence of $N_\textrm{ox}$ on $V_\textrm{th,iso}$ for test diodes with a floating guard-ring. The measured $C_\textrm{geom}$ 
    of the `Half'-diode 
    is scaled to the $C_\textrm{geom}$ of the `Large'-diodes. The three measured values of $V_\textrm{th,iso}$ are reproduced by the simulation with $N_\textrm{ox}$-values of $1.0\times10^{11}$, $1.18\times10^{11}$ and $1.245\times10^{11}~\textrm{cm}^{-2}$. (b) Results for a test diode with guard-ring grounded or floating and with $V_\textrm{th,iso}(195~\textrm{V})<V_\textrm{fd}(280~\textrm{V})$. The simulation applied $N_\textrm{ox}=1.0\times10^{11}~\textrm{cm}^{-2}$.
}
\label{Vth_meas_TCAD_s1L}
\end{figure}

With the approach presented in~\ref{App1}, $V_\textrm{fd}=280~\textrm{V}$ was extracted from the $C_\textrm{diode}$-curve in Figure~\ref{TD_gnd_float} and the derived value of $N_\textrm{B}=4.0\times10^{12}~\textrm{cm}^{-3}$ (utilizing the relation $N_\textrm{B}=2\epsilon_\textrm{s}V_\textrm{fd}/(ed^2)$, where $e$ is the elementary charge \cite{Sze1981}) was applied as an input to the 
test-diode simulation. 
As displayed in Figure~\ref{TD_gnd_float}, by tuning the input $N_\textrm{ox}$ both the voltage of $V_\textrm{th,iso}$ and the test-diode $CV$-characteristics 
with its guard-ring either grounded or floating 
can be closely reproduced by the simulation.
%
%
From further measurements and simulations with varied input $N_\textrm{ox}$ focusing on the $CV$-curves of a test diode with its guard-ring floating in Figure~\ref{Vth_measTCAD}, it is evident that $V_\textrm{th,iso}$ is not dependent on $V_\textrm{fd}$.

%
In Figure~\ref{s1L_gnd_float}, where $V_\textrm{th,iso}{\approx}V_\textrm{fd}-85~\textrm{V}$, the influence of the two factors 
in the evolution of the active volume (involved in the test-diode pad $CV$-measurement) with reverse bias voltage is visible:
\begin{itemize}
\item Isolation between the test-diode pad and the guard-ring.
\item The presence 
or absence 
of opposing electric field-lines from the guard-ring, corresponding to whether the guard-ring is grounded or floating, respectively.
\end{itemize}
The sensitivity to $V_\textrm{th,iso}$ is also visible in the $CV$-measurement with the guard-ring grounded, as the slight non-linearity in the dynamic-region of the $C^{-2}V$-curve vanishes above $V_\textrm{th,iso}$. This suggests that the conduction-channel is also present when the guard-ring is grounded, keeping the test-diode pad and the guard-ring shorted at $V<V_\textrm{th,iso}$, and indicating that $V_\textrm{th,iso}$ is independent of the two $CV$-measurement configurations in Figure~\ref{s1L_gnd_float}. The opposing field-lines from the grounded guard-ring stop the lateral expansion of the pad's active volume with bias voltage, which mitigates the influence of the shorted electrodes on the $C^{-2}V$-curve at voltages below $V_\textrm{th,iso}$. The same 
feature is also visible throughout the dynamic-region of the measured $C^{-2}V$-curve with grounded guard-ring in Figure~\ref{TD_gnd_float}, where $V_\textrm{th,iso}$ is about 5 V below $V_\textrm{fd}$. Additionally, both figures display that the aforementioned feature in the measured $CV$-characteristics of the test diode with grounded guard-ring is 
not reproduced by the 2D-simulation. This is likely to be due to the limitation of the cross-sectional 2D-structure to reproduce the subtle effect from the lateral expansion of the guard-ring field lines that confine the active volume of the test-diode pad from its four sides at the voltages where the pad and the guard-ring are still shorted. Thus, due to considerably more pronounced influence from $V_\textrm{th,iso}$ on the $CV$-characteristics of a test diode with floating guard-ring, the combined experimental and simulation investigation of the correlation between $V_\textrm{th,iso}$ and field-region $N_\textrm{ox}$ is focused on 
this configuration.

Finally, the absence of opposing field-lines from the guard-ring in the $CV$-measurements of the test diode with floating guard-ring in Figures~\ref{TD_gnd_float} and~\ref{Vth_meas_TCAD_s1L} enables greater lateral expansion of the active volume of the test-diode pad even when it is isolated from the guard-ring ($V>V_\textrm{th,iso}$), resulting in higher $C_\textrm{geom}$ (with $A=A_\textrm{diode}+A_\textrm{lateral}$ in Eq.~\ref{eq1}) than the measurement with the guard-ring grounded.
%
%

\begin{figure}[htb!]
     \centering
     \subfloat[]{\includegraphics[width=.5\textwidth]{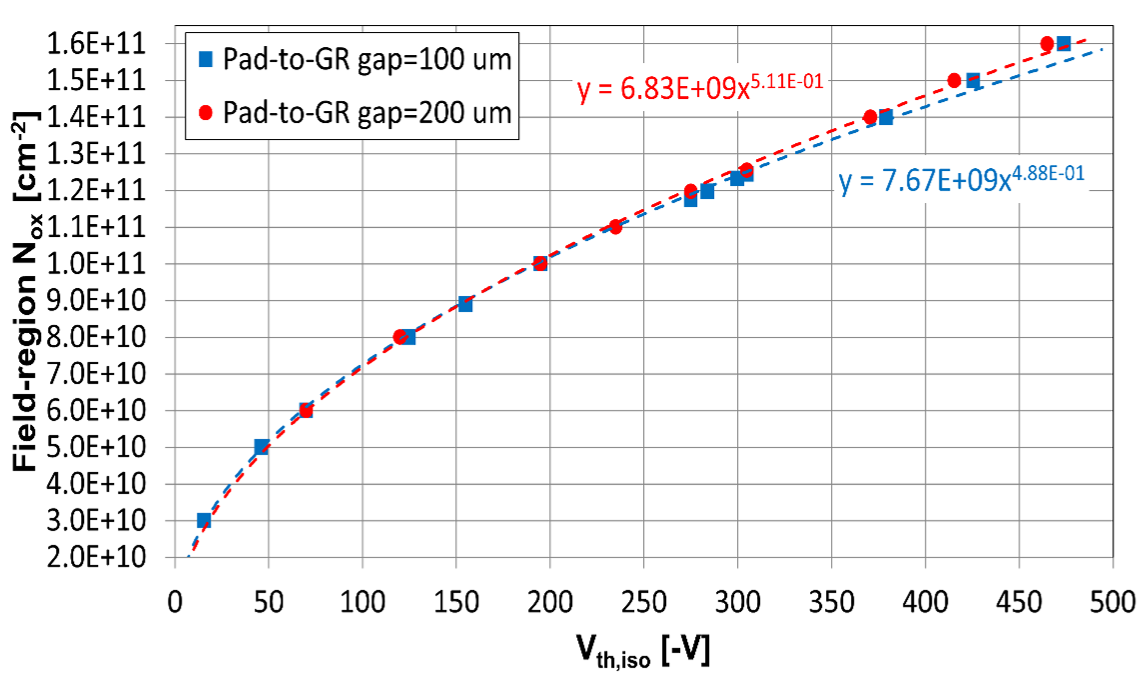}\label{TD_NoxVsVth}}\\
     \subfloat[Sample 1.]{\includegraphics[width=.495\textwidth]{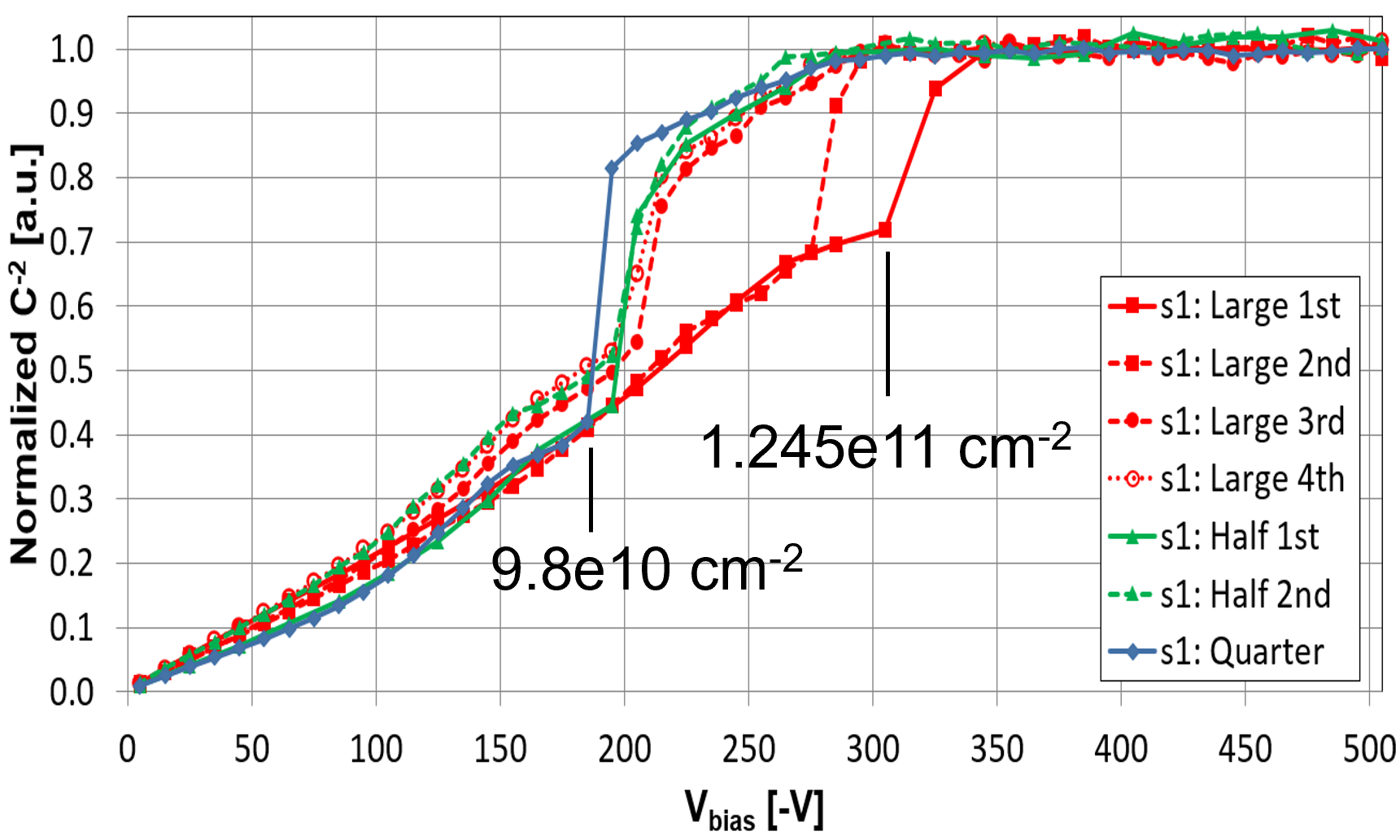}\label{s1_LHQ}}\hspace{1mm}%
     \subfloat[Samples 2 and 3.]{\includegraphics[width=.492\textwidth]{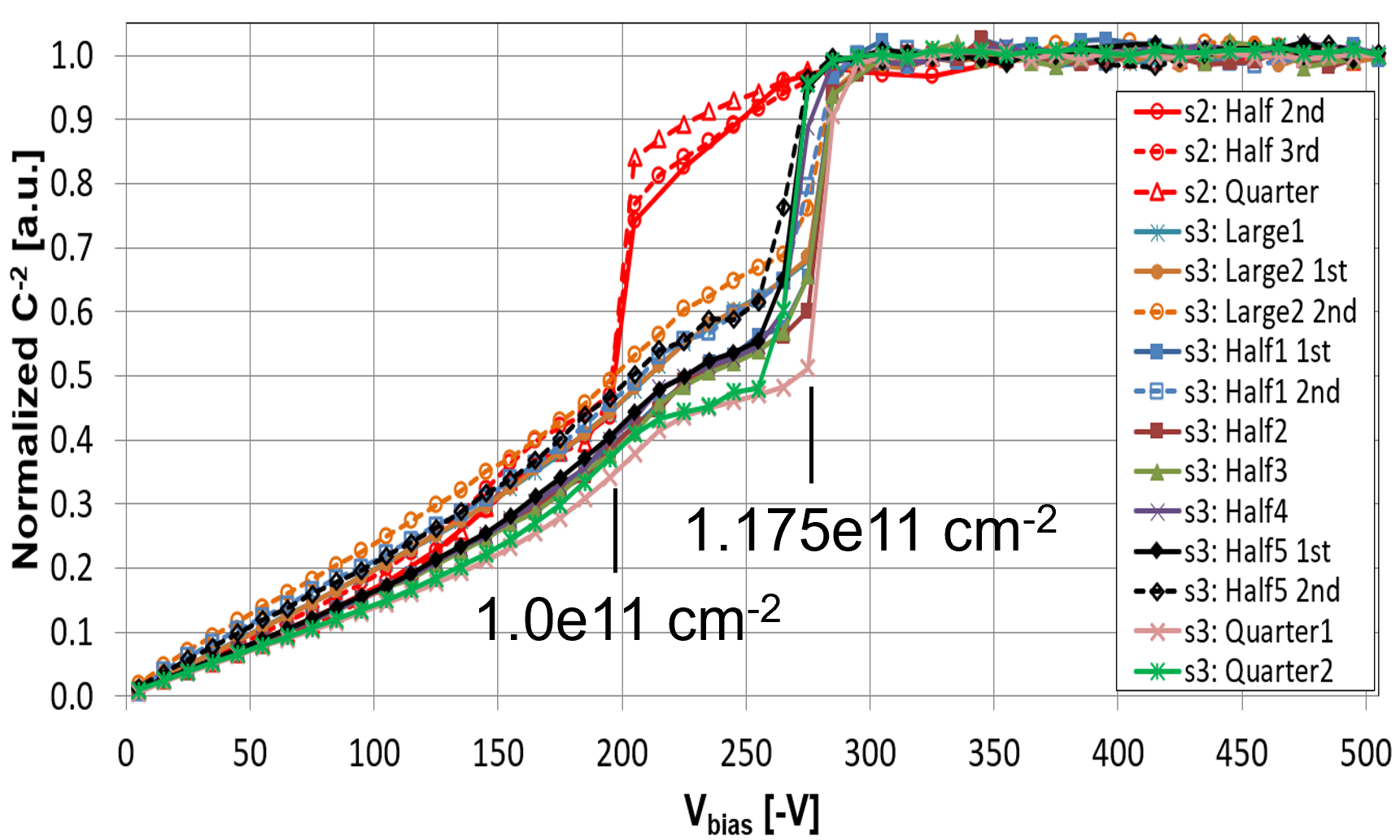}\label{s2s3_LHQ}}
    \caption{\small (a) The simulated dependence of the field-region $N_\textrm{ox}$ on $V_\textrm{th,iso}$ in a 300-$\upmu\textrm{m}$-thick $n$-on-$p$ test diode for two gap widths between the test-diode pad and its floating guard-ring (`Pad-to-GR gap'). The fit parameters for the two data sets are displayed in the plot. (b) and (c) Measured $CV$-characteristics of 14 300-$\upmu\textrm{m}$-thick $n$-on-$p$ test diodes with 
    guard-ring floating. The three `half-moon' samples including the test diodes are named as `s1--3', while repeated measurements on a test diode are indicated as `1st--4th'. The extreme values of field-region $N_\textrm{ox}$, extracted fron the simulations utilizing the fit for 100-$\upmu\textrm{m}$ pad-to-GR gap in Figure~\ref{TD_NoxVsVth}, 
    are indicated. 
}
\label{s1s2s3}
\end{figure}

With additional simulated data-points 
Figure~\ref{TD_NoxVsVth} displays a power-law dependence between $N_\textrm{ox}$ and $V_\textrm{th,iso}$ by the fit $N_\textrm{ox}=7.67\times10^{9}{|V_\textrm{th,iso}|}^{0.488}$ for the 100-$\upmu\textrm{m}$-width pad-to-guard-ring gap of the measured test diodes.  
Doubled pad-to-guard-ring gap in Figure~\ref{TD_NoxVsVth} shows a weak dependence of $V_\textrm{th,iso}$ on the gap width, which becomes negligible at $N_\textrm{ox}<1.2\times10^{11}~\textrm{cm}^{-2}$.
Thus, Figure~\ref{TD_NoxVsVth} provides a tool to determine field-region $N_\textrm{ox}$ from the measured $V_\textrm{th,iso}$ of the $CV$-curve of an $n$-on-$p$ test diode with a floating guard-ring 
or more generally, in the field-region 
between two $n^+$-electrodes with no isolation implant.

Figures~\ref{s1_LHQ} and~\ref{s2s3_LHQ} present the measured $CV$-characteristics of 14 test diodes with guard-ring floating from three `half-moon' samples diced off from the corners of one 6-inch wafer. As shown in Figure~\ref{Nox_TD_MOS}, the values of $V_\textrm{th,iso}$ vary between the `half-moon' samples and with the position on the sample, indicating a non-uniformity of $N_\textrm{ox}$ over the 6-inch wafer, while no dependence on the test-diode size is visible. 
One of the test diodes (sample 1 `Large' in Figure~\ref{s1_LHQ}) displays also a significant sensitivity to repeated measurements with ${\Delta}V_\textrm{th,iso}\approx100~\textrm{V}$ between the first and fourth measurements, after which no change of $V_\textrm{th,iso}$ was observed with further measurements. This could indicate a suppression of electrons pulled to the Si/SiO$_2$-interface by the positive oxide charge, but since repeated measurements on other test diodes showed negligible effect on $V_\textrm{th,iso}$, understanding the nature of this effect is left out of the scope of the following analysis.

Applying the 
fit from Figure~\ref{TD_NoxVsVth} to all measured $V_\textrm{th,iso}$ in Figures~\ref{s1_LHQ} and~\ref{s2s3_LHQ} gives for field-region $\overline{N}_\textrm{ox}=(1.10\pm0.09)\times10^{11}~\textrm{cm}^{-2}$, while the measurements of the MOS-capacitors on samples 1 and 2 (sample 3 with nine test-diodes shown in the bottom-part of Figure~\ref{TD_MOS_6in} did not contain any MOS-capacitors) result in gate $\overline{N}_\textrm{ox}=(1.05\pm0.09)\times10^{11}~\textrm{cm}^{-2}$. Thus, the measurement-simulation and direct-measurement approaches to extract the field-region and the gate $\overline{N}_\textrm{ox}$, respectively, 
are in agreement within uncertainty. Figure~\ref{Nox_TD_MOS} displays a comparison between the individual $N_\textrm{ox}$-results extracted from the field-region of the test diodes and from the MOS-capacitor measurements. 
%
%
\begin{figure}[htb!]
     \centering
    \includegraphics[width=.62\textwidth]{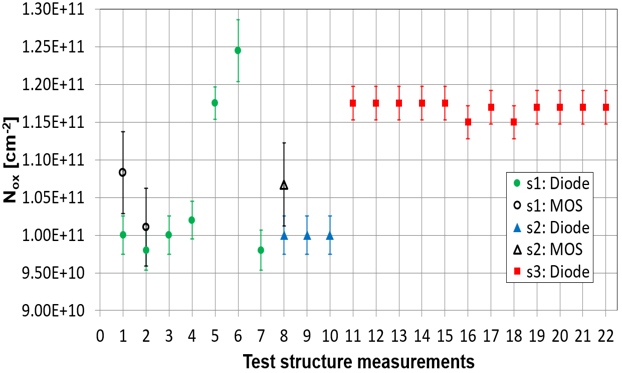}\label{Nox_TD_MOS}
    \caption{\small A comparison of the field-region $N_\textrm{ox}$ extracted from the test diode $V_\textrm{th,iso}$-values in Figures~\ref{s1_LHQ} and~\ref{s2s3_LHQ} between the three `half-moon' samples, as well as with gate $N_\textrm{ox}$ extracted from the $V_\textrm{fb}$-values of the MOS-capacitors on the samples 1 and 2 (`s1--2').
}
\label{Nox_TD_MOS}
\end{figure}
%

\begin{figure}[htb!]
     \centering
     \subfloat[]{\includegraphics[width=.495\textwidth]{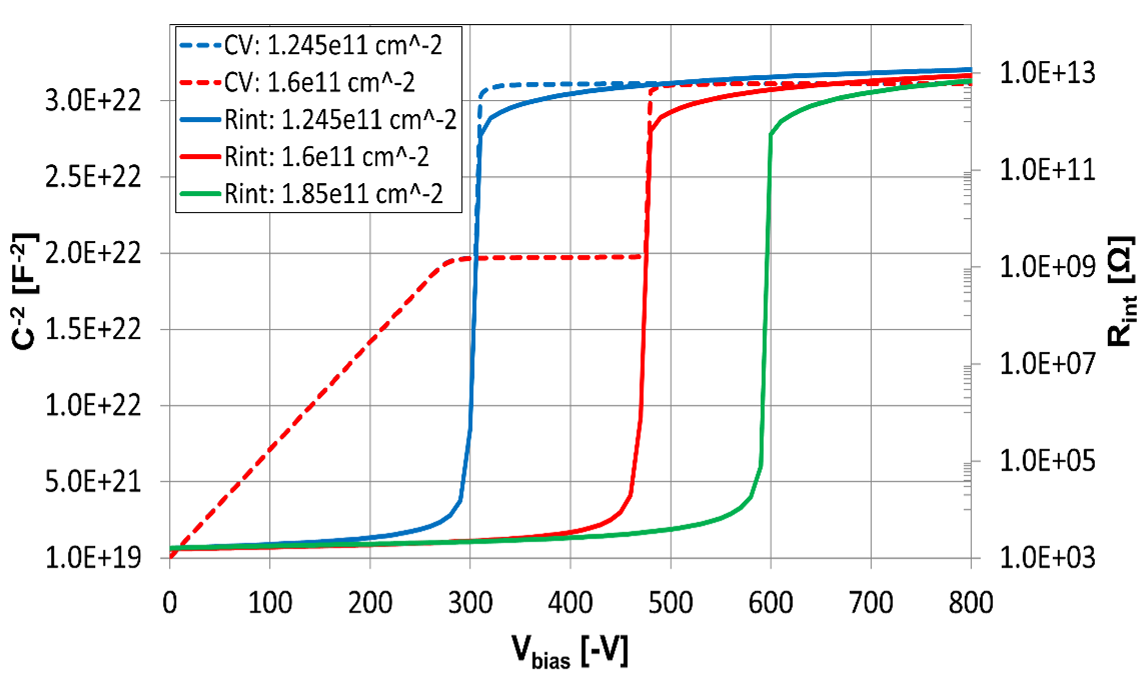}\label{TD_Rint}}\hspace{1mm}%
     \subfloat[]{\includegraphics[width=.495\textwidth]{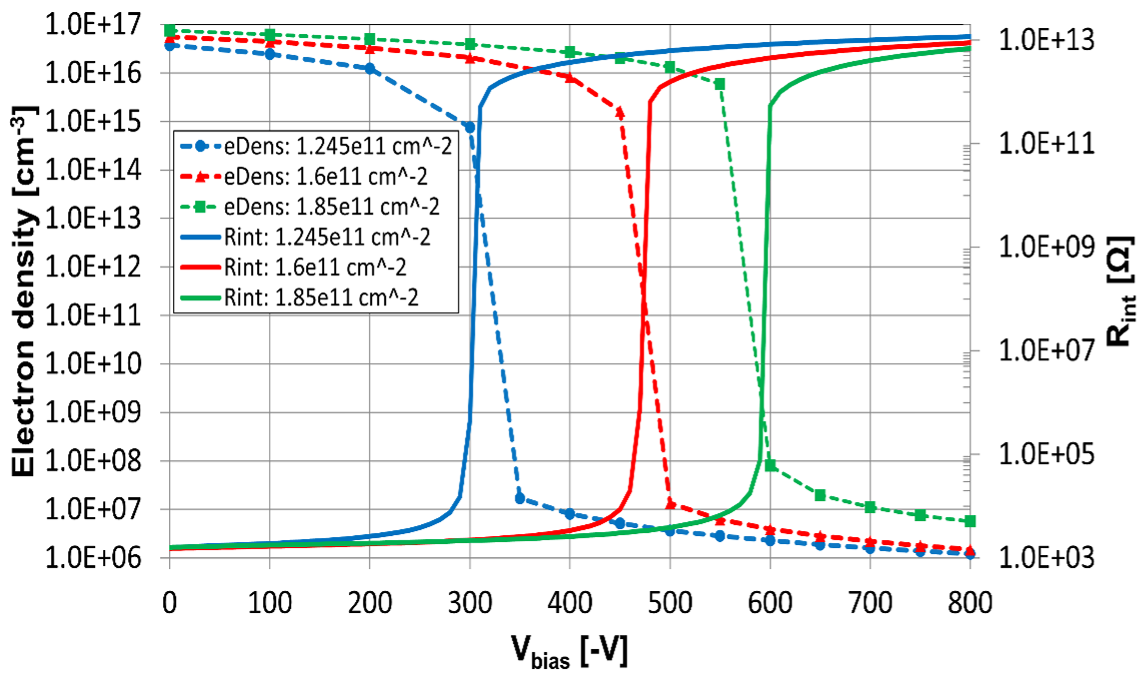}\label{TD_Rint_eDens}}
    \caption{\small Comparison of 
simulated $V_\textrm{th,iso}$ in the 
$R_\textrm{int}$-characteristics of a 300-$\upmu\textrm{m}$-thick $n$-on-$p$ test diode for two $N_\textrm{ox}$-levels from Figure~\ref{Vth_measTCAD}, 
along with $N_\textrm{ox}$-level that results in $V_\textrm{th,iso}(R_\textrm{int})=600~\textrm{V}$ with (a) $V_\textrm{th,iso}$ in the $CV$-characteristics of the two lower $N_\textrm{ox}$-levels and (b) the change in electron density at 1 nm depth from the Si/SiO$_2$-interface in the middle of the inter-electrode gap.
}
\label{TD_Rint_CV_eDens}
\end{figure}

The correlation of $V_\textrm{th,iso}$ observed in $CV$-characteristics to the change in the inter-electrode resistance ($R_\textrm{int}$) for $n^+$-electrodes without isolation implant is presented in Figure~\ref{TD_Rint}. 
It is evident that the $CV$-extracted $V_\textrm{th,iso}$ 
is identical to the voltage where $R_\textrm{int}$ increases abruptly by about ten-orders-of-magnitude, i.e., $V_\textrm{th,iso}(CV)=V_\textrm{th,iso}(R_\textrm{int})$. The simulated change 
of $R_\textrm{int}$ above $V_\textrm{th,iso}$ 
is also in close agreement with experimental results \cite{Gosewich2021_J}. 
Hence, the electron removal from the inter-electrode gap by applied reverse bias voltage in Figures~\ref{eD_PadToGR} and~\ref{TD_Rint_eDens} results in a high level of $R_\textrm{int}$ which isolates the $n^+$-electrodes at a given $V_\textrm{th,iso}$ that can be observed in a $CV$-measurement. Since in $R_\textrm{int}$-simulation both $n^+$-electrodes are biased (unlike in $CV$-simulation where the guard-ring is floating), Figure~\ref{TD_Rint} also displays that $V_\textrm{th,iso}$ is independent of whether one or both of the electrodes are biased. 
This supports the observation of the measured $C^{-2}V$-curves with either grounded or floating guard-ring in Figure~\ref{s1L_gnd_float} where both configurations show sensitivity to the same $V_\textrm{th,iso}$. Finally, the $R_\textrm{int}$-curve for the highest $N_\textrm{ox}$ in Figure~\ref{TD_Rint} shows that for $N_\textrm{ox}<1.85\times10^{11}~\textrm{cm}^{-2}$, $n^+$-electrodes of a 300-$\upmu\textrm{m}$-thick $n$-on-$p$ test diode, i.e., a pad-sensor, 
remain isolated without an isolation implant when operated at the bias voltage of 600 V.
%
%

\begin{figure}[htb!]
     \centering
     \subfloat[]{\includegraphics[width=.495\textwidth]{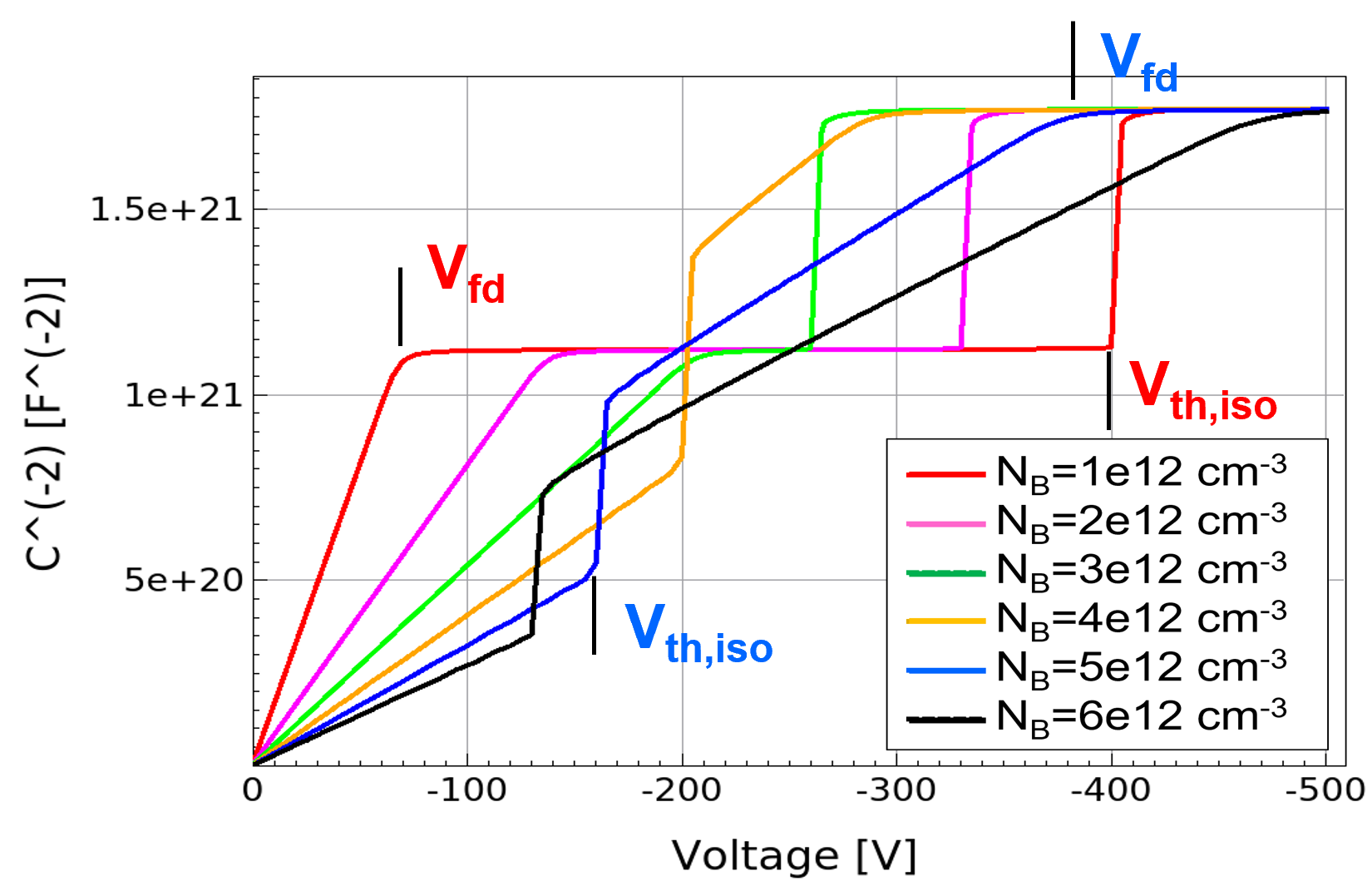}\label{Nb_Vth}}\hspace{1mm}%
     \subfloat[]{\includegraphics[width=.495\textwidth]{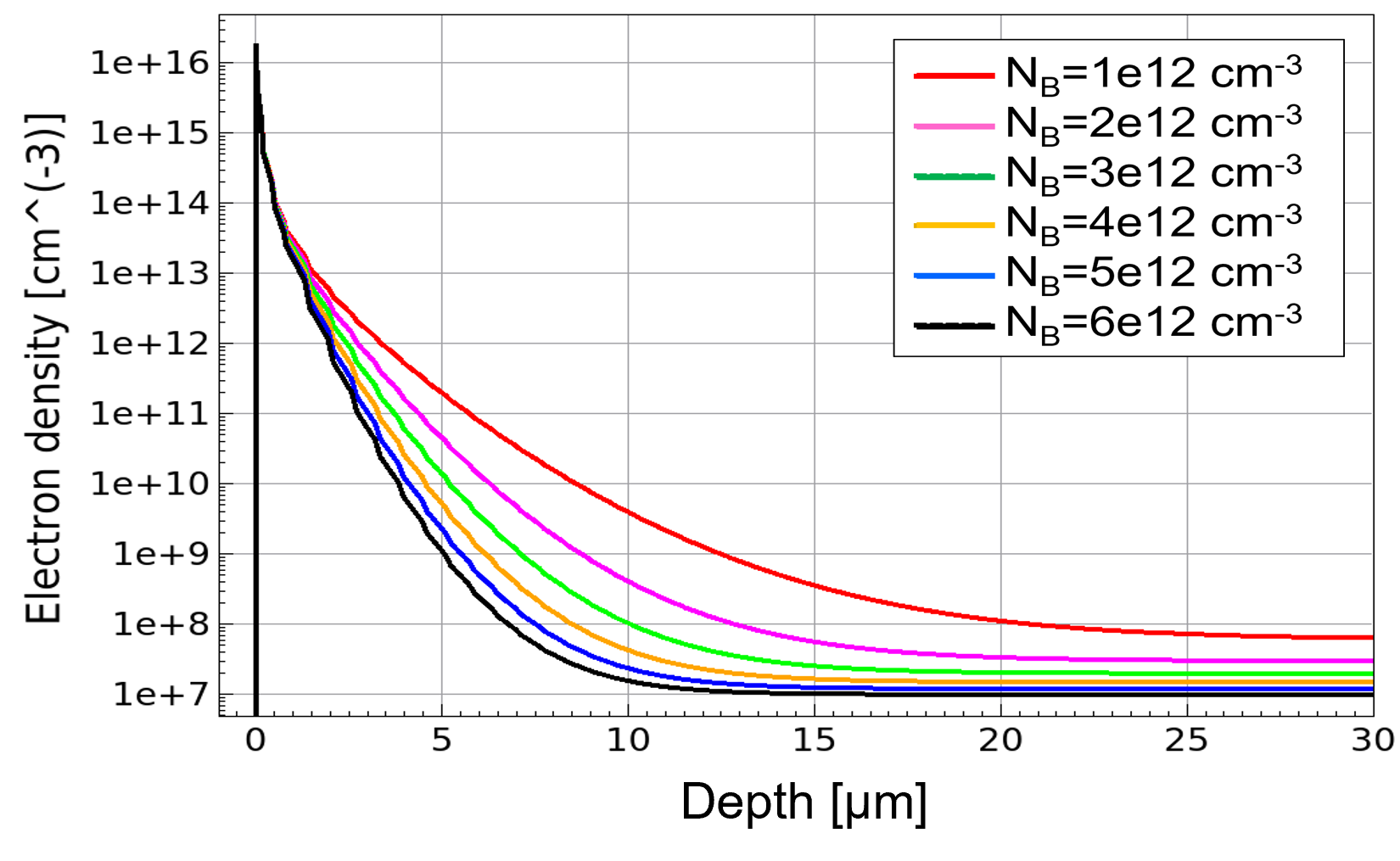}\label{Nb_eD}}\\
     \subfloat[]{\includegraphics[width=.495\textwidth]{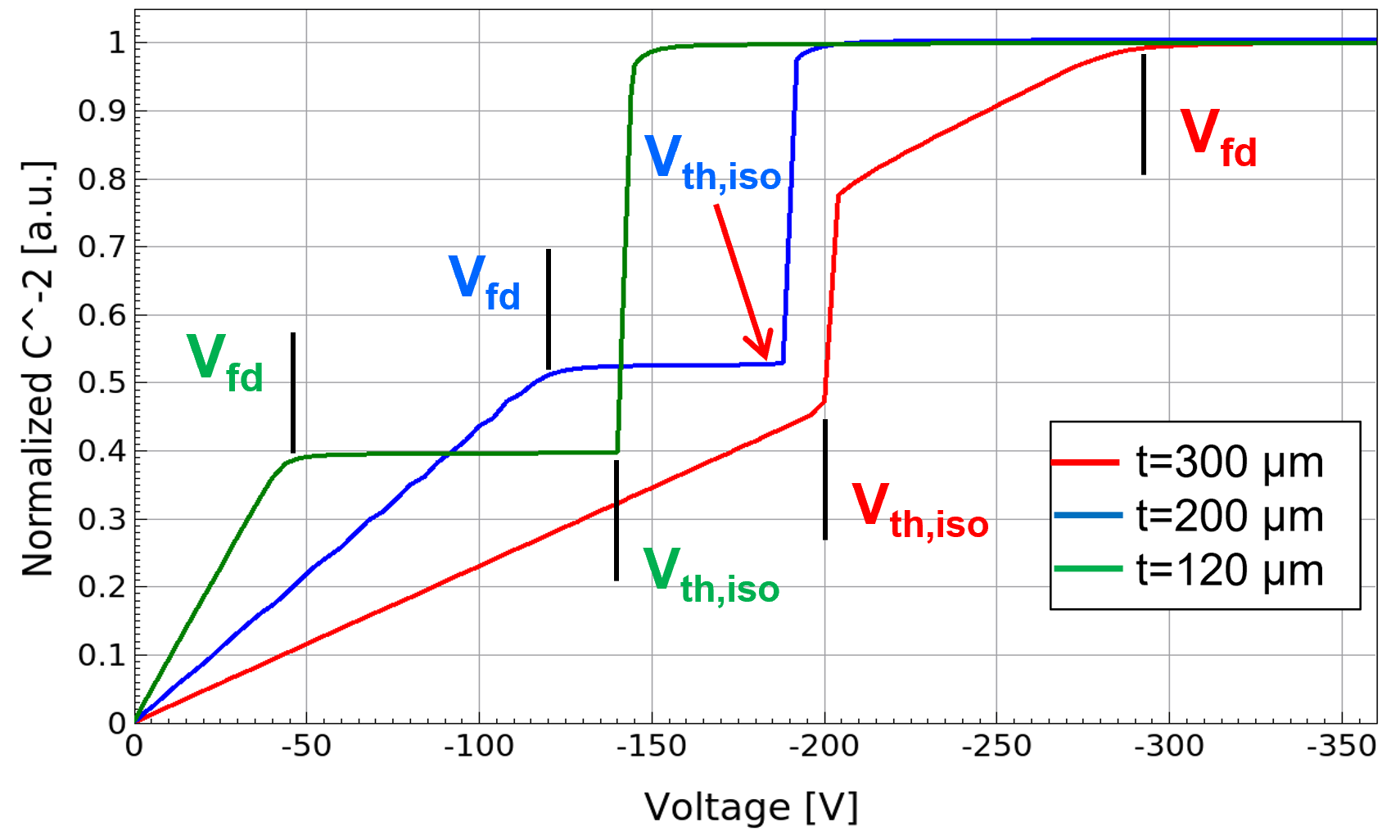}\label{t_dep}}
    \caption{\small Simulation results of an $n$-on-$p$ test-diode with floating guard-ring for the influence of $N_\textrm{B}$ and 
    $t$ on $V_\textrm{th,iso}$. (a) Change of $V_\textrm{th,iso}$ with $N_\textrm{B}$ for constant $t=300~\upmu\textrm{m}$ and $N_\textrm{ox}=1\times10^{11}~\textrm{cm}^{-2}$. (b) Corresponding electron density to Figure~\ref{Nb_Vth} close to the Si/SiO$_2$-interface at mid-gap between pad and guard-ring for $V=0$. (c) Change of $V_\textrm{th,iso}$ with $t$ for constant $N_\textrm{B}=4\times10^{12}~\textrm{cm}^{-3}$ and $N_\textrm{ox}=1\times10^{11}~\textrm{cm}^{-2}$. 
}
\label{Nb_thickness}
\end{figure}

The sensitivity of $V_\textrm{th,iso}$ to the influence of $N_\textrm{B}$ and the active thickness ($t$) of the sensor is presented in Figure~\ref{Nb_thickness}. With a constant $t$ and $N_\textrm{ox}$, 
$V_\textrm{th,iso}$ displays a strong dependence on $N_\textrm{B}$ in Figure~\ref{Nb_Vth} with $V_\textrm{th,iso}>V_\textrm{fd}$ at $N_\textrm{B}\leq3\times10^{12}~\textrm{cm}^{-3}$ and $V_\textrm{th,iso}<V_\textrm{fd}$ at $N_\textrm{B}\geq4\times10^{12}~\textrm{cm}^{-3}$. As shown in Figure~\ref{Nb_eD}, increased $N_\textrm{B}$ suppresses the minority carrier density close to Si/SiO$_2$-interface leading to reduced voltages required to
isolate the electrodes, thus moving $V_\textrm{th,iso}$ to lower voltages in Figure~\ref{Nb_Vth}. Additionally, with a constant $N_\textrm{B}$ and $N_\textrm{ox}$, $V_\textrm{th,iso}$ displays significant dependence on $t$ in Figure~\ref{t_dep}, with $V_\textrm{th,iso}>V_\textrm{fd}$ at $t<300~\upmu\textrm{m}$ and ${\Delta}V_\textrm{th,iso}{\approx}-60~\textrm{V}$ when $t$ is reduced from 300 to 120~$\upmu\textrm{m}$. Since for a given voltage higher electric fields are generated in a thinner sensor, a lower voltage for a thinner sensor is required to generate sufficient 
field to reduce electron density in the inter-electrode gap to a level that isolates the electrodes, displayed by lower $V_\textrm{th,iso}$ with reduced $t$ in Figure~\ref{t_dep}. As a consequence, $CV$-extracted $V_\textrm{th,iso}$($N_\textrm{ox},N_\textrm{B},t$) and therefore the results in Figures~\ref{TD_NoxVsVth} and~\ref{TD_Rint} should be considered for fixed $N_\textrm{B}=4.0\times10^{12}~\textrm{cm}^{-3}$ and $t=300~\upmu\textrm{m}$.
%
\subsection{8-inch HGCAL sensors}
\label{HGCALsensors}
To extend the investigation presented in Section~\ref{TDs}
to 8-inch HGCAL sensors, the region between the two guard-rings in the periphery of a HGCAL sensor displayed in Figure~\ref{HGCALedge} was considered. To limit the lateral extension of the active region of the multi-channel sensor toward the physical wafer edge, the guard-ring closer to the HV-ring (`Floating GR' in Figure~\ref{HGCALedge}) is left floating in the ARRAY setup's probe card by default. The electrical configuration between the biased (bias-GR) and floating guard-rings in the HGCAL sensor is then identical to the test diode with a floating guard-ring (i.e., no isolation implant between the biased and floating $n^+$-electrodes). The main difference between the devices comes from their geometry, since instead of a square-shaped pad of the test diode, the biased electrode is now a $115~\upmu\textrm{m}$-wide strip with a hexagon shape for the full sensors in Figures~\ref{LDfull} and~\ref{HDfull}, and polygon and rectangle shapes for the Multi-Geometry Wafer (MGW)-sensor in Figure~\ref{HDmgw}. Additionally, the gap between the 
two electrodes is 62 and $100~\upmu\textrm{m}$ for the HGCAL sensor and test diode, respectively. Thus, despite the differences in the device geometry to the test diodes the $CV$-characteristics of the HGCAL sensors' bias-GR are expected to 
provide an observation of $V_\textrm{th,iso}$. 

\begin{figure}[htb!]
     \centering
     \subfloat[]{\includegraphics[width=.495\textwidth]{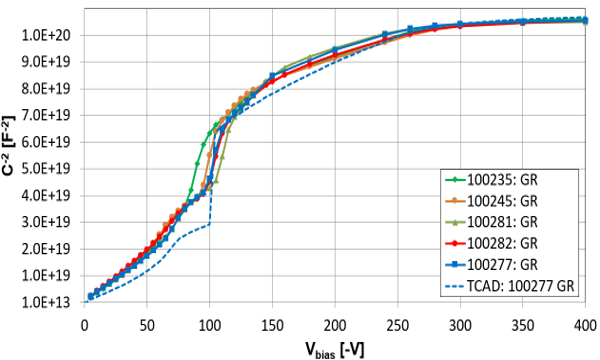}\label{GR_CV_300um_full_oxC}}\hspace{1mm}%
      \subfloat[]{\includegraphics[width=.495\textwidth]{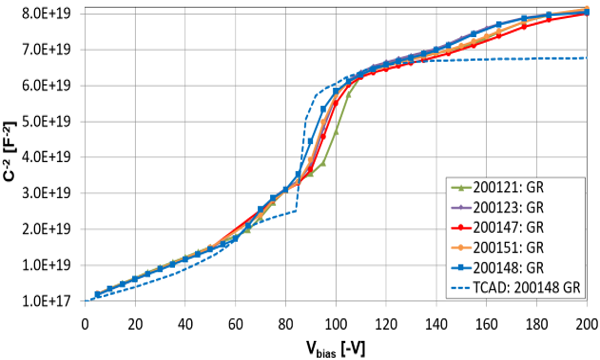}\label{GR_CV_200um_full_oxC}}\\
      \subfloat[]{\includegraphics[width=.495\textwidth]{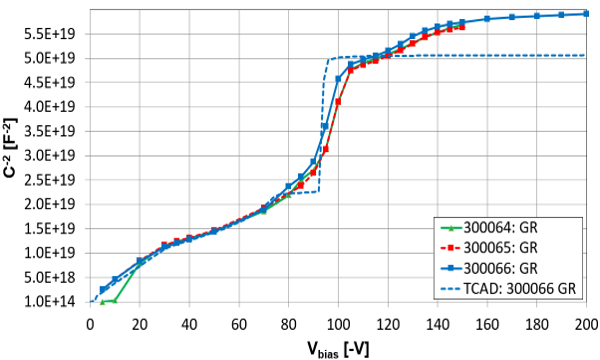}\label{GR_CV_120um_full_oxC}}
    \caption{\small Measured and simulated $CV$-characteristics for 
    the biased guard-ring (`GR') in full 8-inch HGCAL sensors with oxide type `C'. 
    Sensor IDs are indicated in the legends, while the simulation input values of $N_\textrm{B}$ are given in Figure~\ref{Vfd_Nb_HGCALfull}. (a) Measured results of five 300-$\upmu\textrm{m}$-thick LD-sensors with $V_\textrm{th,iso}$ of the sensor `100277' reproduced by the simulation with the input $N_\textrm{ox}=6.90\times10^{10}~\textrm{cm}^{-2}$. Out of the 20 measured sensors, only $C^{-2}V$-curves with significant variation in $V_\textrm{th,iso}$ are included in the plot for clarity. (b) Measured results of five 200-$\upmu$m-thick LD-sensors with $V_\textrm{th,iso}$ of the sensor `200148' reproduced by the simulation with the input $N_\textrm{ox}=6.45\times10^{10}~\textrm{cm}^{-2}$. (c) Measured results of three 120-$\upmu$m-thick HD-sensors with $V_\textrm{th,iso}$ of the sensor `300066' reproduced by the simulation with the input $N_\textrm{ox}=6.65\times10^{10}~\textrm{cm}^{-2}$. 
}
\label{GR_CV_HGCALfull}
\end{figure}

\begin{figure}[htb!]
     \centering
     \subfloat[]{\includegraphics[width=.495\textwidth]{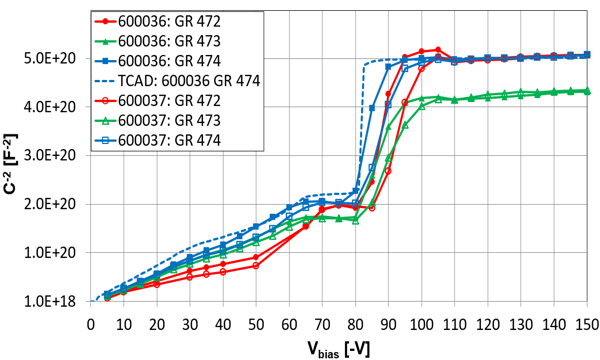}\label{GR_CV_120um_MGW_oxB}}\hspace{1mm}%
     \subfloat[]{\includegraphics[width=.495\textwidth]{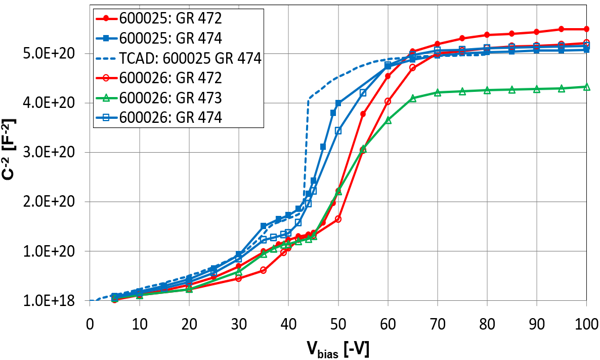}\label{GR_CV_120um_MGW_oxD}}
    \caption{\small Measured and simulated $CV$-characteristics for 
    the biased guard-rings (`GR 472--474'corresponding to the channel numbers in Figure~\ref{HDmgw}) in 8-inch partial (`bottom-cut') 120-$\upmu$m HD HGCAL sensors. The simulation input values of $N_\textrm{B}$ are given in Figure~\ref{Vfd_Nb_HGCAL_MGW}. (a) Measured results of two sensors with oxide type `B' with $V_\textrm{th,iso}$ of the `GR 474' of the sensor `600036' reproduced by the simulation with the input $N_\textrm{ox}=5.98\times10^{10}~\textrm{cm}^{-2}$. (b) Measured results of two sensors with oxide type `D' with $V_\textrm{th,iso}$ of the `GR 474' of the sensor `600025' reproduced by the simulation with the input $N_\textrm{ox}=5.15\times10^{10}~\textrm{cm}^{-2}$. 
}
\label{GR_CV_HGCAL_MGW}
\end{figure}

As 
shown in Figures~\ref{GR_CV_HGCALfull} and~\ref{GR_CV_HGCAL_MGW}, 
bias-GR $CV$-characteristics of all measured HGCAL sensor types display $V_\textrm{th,iso}$ either below (Figures~\ref{GR_CV_300um_full_oxC},~\ref{GR_CV_200um_full_oxC} and~\ref{GR_CV_120um_MGW_oxD}) or above (Figures~\ref{GR_CV_120um_full_oxC} and~\ref{GR_CV_120um_MGW_oxB}) $V_\textrm{fd}$ (values of $\overline{V}_\textrm{fd}$ are given in~\ref{App1}). In all cases the $n^+$-electrodes without $p$-stop isolation implant become isolated at $V_\textrm{th,iso}\leq100~\textrm{V}$, which is significantly below the 
operating voltage of 600 V considered for HGCAL \cite{Phase2}. The wafer-position dependence of $V_\textrm{th,iso}$ observed in the test-diode results in Figures~\ref{s1_LHQ} and~\ref{s2s3_LHQ} is also visible in Figure~\ref{GR_CV_HGCAL_MGW} where the `bottom-cut' partial sensors have three bias-GRs enclosing three regions of the sensor, as shown in Figure~\ref{HDmgw}. The different lengths between the rectangular bias-GR at the center of the sensor (channel 473) and the two polygonal bias-GRs adjacent to it (channels 472 and 474) in the `bottom-cut' sensor 
are reflected in the different values of $C_\textrm{geom}$, 
according to Eq.~\ref{eq1}. 

With the observations from Figure~\ref{Nb_thickness}, dedicated $t$ and $N_\textrm{B}$ were applied to each of the modeled sensor structures to reproduce the measured $V_\textrm{th,iso}$ by simulation. 
After $N_\textrm{ox}$-tuning the main disagreement between simulation and measurement for 300-$\upmu\textrm{m}$-thick full sensors in Figure~\ref{GR_CV_300um_full_oxC} is in the region $V<V_\textrm{th,iso}$, where the lower slope of the simulated $C^{-2}V$-curve indicates a wider lateral expansion of the depletion volume before the biased and floating guard-rings become isolated than in the real sensor. In Figure~\ref{GR_CV_200um_full_oxC} the disagreement between simulation and measurement for 200-$\upmu\textrm{m}$-thick full sensors is now in the region $V>V_\textrm{th,iso}$, where the further limited decrease of the active volume involved in the bias-GR $CV$-measurement at $V>V_\textrm{fd}\approx115~\textrm{V}$ is not reproduced by the simulation. The same trend is visible in Figure~\ref{GR_CV_120um_full_oxC} for 120-$\upmu\textrm{m}$-thick full sensors even though 
$V_\textrm{fd}\approx31~\textrm{V}$, thus the feature is only dependent on $V_\textrm{th,iso}$ and present only in the bias-GR $CV$-characteristics of full 200- and 120-$\upmu\textrm{m}$-thick sensors. The onset of the feature has a dependence on active thickness, being about 20 V lower for 120-$\upmu\textrm{m}$-thick sensors than for 200-$\upmu\textrm{m}$ thickness, while its interpretation is left out of the scope of this study.
The sole significant disagreement between simulation and measurement for the partial `bottom-cut' 120-$\upmu\textrm{m}$-thick sensors in Figures~\ref{GR_CV_120um_MGW_oxB} and~\ref{GR_CV_120um_MGW_oxD} is in the abruptness of the decrease of $C$ immediately after $V_\textrm{th,iso}$, indicating that the electrons are being swept from the inter-electrode gap to the positively biased guard-ring more swiftly in the simulation than in measurement.
Considering that the length-ratio of the bias-GRs in the 
`bottom-cut' partial and the full sensors is roughly $\frac{1}{6}$, the results suggest that the 2D TCAD-simulation is able to reproduce more closely the measured $CV$-characteristics of the bias-GRs that enclose a smaller sensor-region. This is also supported by the close agreements between measurements and simulations of test diodes in Section~\ref{TDs} with substantially smaller area and less complex geometry to the bias-GRs of HGCAL sensors.
%
%
%

The investigated HGCAL sensors included three oxide-quality types:`C' for the full sensors, and `B' and `D' for the partial sensors indicating HPK's oxide candidates with the oxide growth thermally and environmentally varied. 
By generating additional data-points for each sensor configuration and oxide type, the extracted $\overline{N}_\textrm{ox}$ are presented in Table~\ref{table_HGCAL}.
For full sensors with oxide type `C' the values of $\overline{N}_\textrm{ox}$ remain within uncertainty for the three thicknesses, while being distinctly higher than $\overline{N}_\textrm{ox}$ of both `B' and `D'. 
Comparison 
with $V_\textrm{fb}$-extracted 
$\overline{N}_\textrm{ox}$ ($\overline{N}_\textrm{ox,gate}$) from the MOS-capacitor measured results in Table~\ref{table_HGCAL} and Figure~\ref{Nox_biasGR_MOS} 
shows matching qualitative trend between the oxide types, while the values of $\overline{N}_\textrm{ox,gate}$ 
remain lower than $\overline{N}_\textrm{ox}$ by a factor $0.77\pm0.13$ (considering average $(6.7\pm0.3)\times10^{10}~\textrm{cm}^{-2}$ of the field-region $\overline{N}_\textrm{ox}$-values for type `C' in Table~\ref{table_HGCAL}) over the three oxide types. 
The disagreement between $\overline{N}_\textrm{ox,gate}$ and $\overline{N}_\textrm{ox}$ observed here is in contrast with the observed corresponding agreement between test structures in Section~\ref{TDs}, which suggests that $N_\textrm{ox}$ in the field region between the biased guard-ring and HV-ring of HGCAL sensors is somewhat higher than the gate $N_\textrm{ox}$ at their respective MOS-capacitors. 
Since the oxide-quality types available for this investigation are only a subset of the multitude of oxide type candidates 
provided by HPK to the HGCAL collaboration, 
further comparisons between the oxide-quality types are 
not included in the scope of this study.
%
\begin{table*}[!t]
\centering
\caption{Active thicknesses, sensor and passivation oxide types, number of measured sensors, 
the mean field-region oxide charge densities ($\overline{N}_\textrm{ox}$), number of measured MOS-capacitors and the mean gate oxide charge densities ($\overline{N}_\textrm{ox,gate}$) at the Si/SiO$_2$-interface. 
}
\label{table_HGCAL}
\begin{tabular}{|c|c|c|c|c|c|c|}
\hline
\multirow{2}{*}{{\bf Active thickness}} & 
\multirow{2}{*}{{\bf {\small Sensor type}}} & \multirow{2}{*}{{\bf {\small Oxide}}} & \multirow{2}{*}{{\bf {\small Measured}}} & \multirow{2}{*}{{\bf $\overline{N}_\textrm{ox}$}} & \multirow{2}{*}{{\bf {\small Measured}}} & \multirow{2}{*}{{\bf $\overline{N}_\textrm{ox,gate}$}}\\[0.9mm]
{\small [$\upmu$m]} & 
{ } & {\bf {\small type}} & {\bf {\small sensors}} & {\small [$\times10^{10}~\textrm{cm}^{-2}$]} & {\bf {\small MOS}} & {\small [$\times10^{10}~\textrm{cm}^{-2}$]}\\
\hline
300 & 
{\small LD full} & C & 20 & $6.74\pm0.14$ & \multirow{3}{*}{6} & \multirow{3}{*}{$5.2\pm0.3$}\\
\cline{1-5}
200 & 
{\small LD full} & C & 5 & $6.64\pm0.12$ &&\\
\cline{1-5}
120 & 
{\small HD full} & C & 4 & $6.8\pm0.2$ &&\\
\hline
120 & 
{\small HD `bottom-cut'} & B & 2 & $6.01\pm0.14$ & 6 & $4.5\pm0.5$\\
\hline
120 & 
{\small HD `bottom-cut'} & D & 2 & $5.32\pm0.17$ & 4 & $4.2\pm0.4$\\
\hline
\end{tabular}
\end{table*}
%
\begin{figure}[htb!]
     \centering
    \includegraphics[width=.62\textwidth]{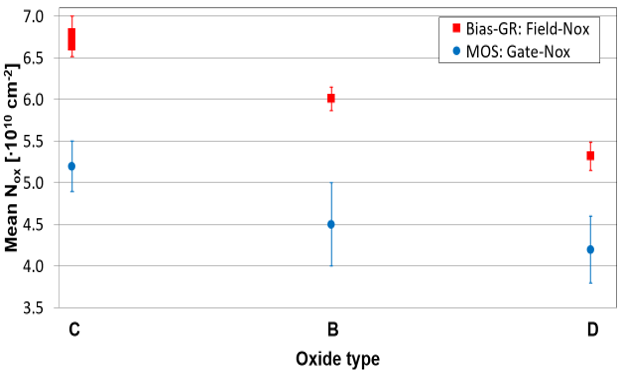}\label{Nox_biasGR_MOS}
    \caption{\small $\overline{N}_\textrm{ox}$-values from Table~\ref{table_HGCAL} for the three oxide quality types, extracted either from bias-GR measurements and simulations (`Field-$N_\textrm{ox}$') or from MOS-capacitor measurements (`Gate-$N_\textrm{ox}$').
%
}
\label{Nox_biasGR_MOS}
\end{figure}
%

\begin{figure}[htb!]
     \centering
     \subfloat[]{\includegraphics[width=.495\textwidth]{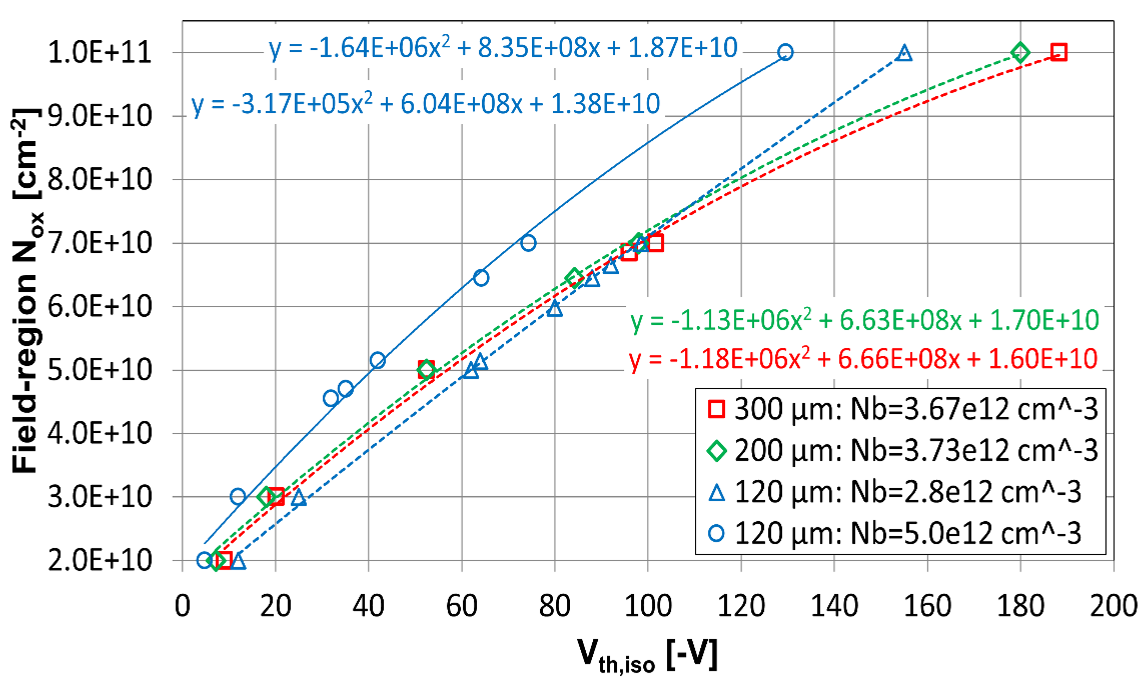}\label{PreS_NoxVth_1e11}}\hspace{1mm}%
     \subfloat[]{\includegraphics[width=.495\textwidth]{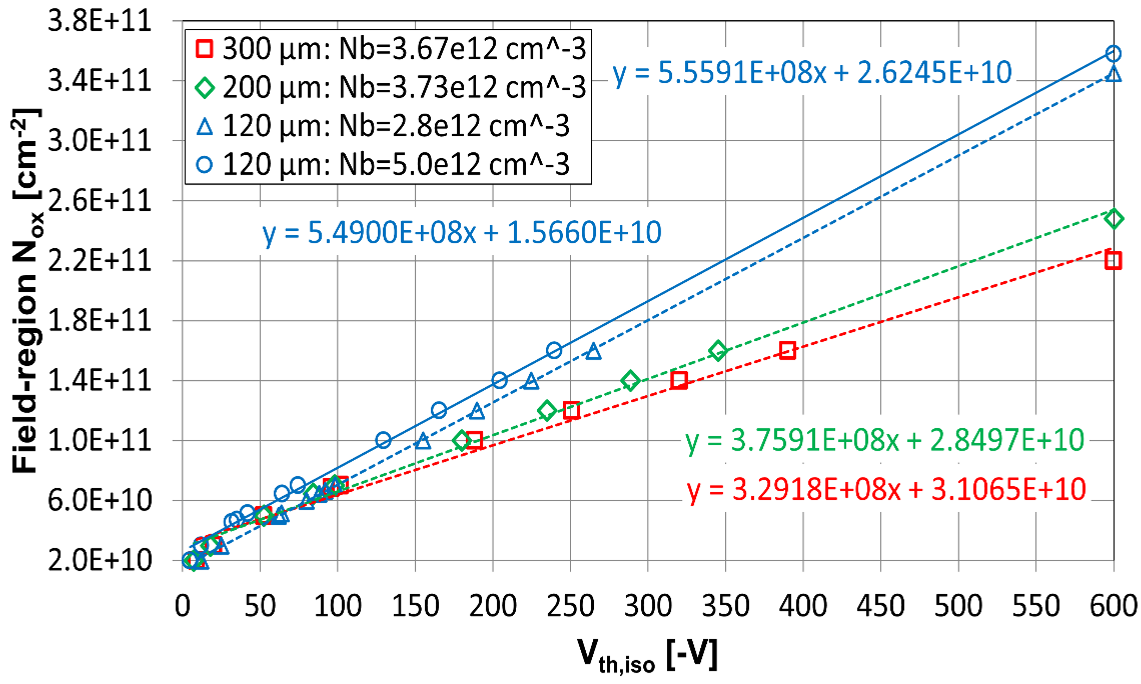}\label{PreS_NoxVth_600V}}
    \caption{\small The simulated dependence of the field-region $N_\textrm{ox}$ on $V_\textrm{th,iso}$ for the three active thicknesses of the 8-inch HGCAL sensors with the input values of $N_\textrm{B}$ indicated in the legends. (a) Simulation results for the input $N_\textrm{ox}\leq1.0\times10^{11}~\textrm{cm}^{-2}$ with second-order polynomial fits to the data-sets. (b) Input $N_\textrm{ox}$-range extended to the levels that result in $V_\textrm{th,iso}=600~\textrm{V}$ with linear fits to the data-sets. The fit parameters are displayed in the plots.
}
\label{PreS_NoxVth}
\end{figure}

Displayed in Figure~\ref{PreS_NoxVth_1e11} the dependence of the field-region $N_\textrm{ox}$ on $V_\textrm{th,iso}$ for the three HGCAL sensor thicknesses (with the simulation input $\overline{N}_\textrm{B}$ from Figures~\ref{Vfd_Nb_HGCALfull} and~\ref{Vfd_Nb_HGCAL_MGW}) can be described by second-order polynomial fits in the region $N_\textrm{ox}\leq1.0\times10^{11}~\textrm{cm}^{-2}$ 
(a range which includes all extracted $\overline{N}_\textrm{ox}$-values in Table~\ref{table_HGCAL}). When the simulated $V_\textrm{th,iso}$-range is extended from $<200~\textrm{V}$ to 600 V, the dependence is best described by linear fits in Figure~\ref{PreS_NoxVth_600V}. The $V_\textrm{th,iso}$-dependence of $N_\textrm{ox}$ at both high and low $N_\textrm{ox}$-regions for HGCAL sensors is then distinct from the power-law dependence of the test diodes in Figure~\ref{TD_NoxVsVth}, suggesting an influence from the substantially different geometries between the simulated device-structures in Sections~\ref{TDs} and~\ref{HGCALsensors}. 
Since the observed $V_\textrm{th,iso}$ contains simultaneously information on $N_\textrm{ox}$ and on the change from low to high levels of $R_\textrm{int}$,
shown in Figure~\ref{TD_Rint}, the simulated data-points in Figure~\ref{PreS_NoxVth_600V} provide threshold-values of $N_\textrm{ox}$ below which $n^+$-electrodes without isolation implants are predicted to be isolated in HGCAL sensors operated at $V_\textrm{bias}=600~\textrm{V}$.
%

%% file: discussion_fT.tex
The results in Section~\ref{HGCALsensors} indicate that pre-irradiated $n$-on-$p$ sensors without isolation implants between the channels can reach electrical isolation, and thus uncompromised position resolution between the channels, by the application of sufficient reverse $V_\textrm{bias}$ (for the $N_\textrm{ox}$-values in Table~\ref{table_HGCAL} at about $V_\textrm{bias}>40-100~\textrm{V}$). As the highest $N_\textrm{ox}$-values in SiO$_2$-passivated Si-sensors with $\langle100\rangle$ crystal orientation are typically well below $2\times10^{11}~\textrm{cm}^{-2}$ \cite{Peltola2023,Peltola2017_J}, also these levels are expected to be within the isolation range of the operation voltage 600 V, as shown in Figure~\ref{PreS_NoxVth_600V} (for $2\times10^{11}~\textrm{cm}^{-2}$ all sensor types become isolated in the range of about 320--500 V). 

In irradiated sensors, the surface damage caused by ionizing radiation from charged particles, X-rays or gammas 
includes the accumulation of $N_\textrm{f}$ and deep donor- and acceptor-type $N_\textrm{it}$ ($N_\textrm{it,don}$ and $N_\textrm{it,acc}$, respectively) 
that contribute to the net $N_\textrm{ox}$ at the Si/SiO$_2$-interface.
Previous experimental $R_\textrm{int}$-results of 200- and 290-$\upmu\textrm{m}$-thick $n$-on-$p$ strip-sensors without $p$-stop implants neutron or proton irradiated to $\Phi=6\times10^{14}~\textrm{n}_\textrm{eq}\textrm{cm}^{-2}$, 
corresponding to the estimated Total Ionizing Doses (TID) of 6 and 870 kGy, respectively,  
show high levels of inter-strip isolation at  $V_\textrm{bias}>100~\textrm{V}$ (see Figure 9 in ref. \cite{Gosewich2021_J}).
As described in ref. \cite{Peltola2023}, 
radiation accumulated levels of $N_\textrm{it,don}$ and $N_\textrm{it,acc}$ comparable to $N_\textrm{f}$ introduce a dynamic characteristic to the net $N_\textrm{ox}$, as the fractions of fully occupied $N_\textrm{it}$ evolve with applied voltage. Contribution from fully occupied $N_\textrm{it,acc}$ results in considerably low levels of net $N_\textrm{ox}$ in the field-region between the positively biased $n^+$-electrodes (see Figure 8 in ref. \cite{Peltola2023}) of a position sensitive $n$-on-$p$ sensor, which remain within the isolation range in Figure~\ref{PreS_NoxVth_600V}. Thus, the radiation-induced accumulation of $N_\textrm{it,acc}$ at the Si/SiO$_2$-interface has a beneficial impact on the inter-electrode isolation in irradiated $n$-on-$p$ sensors, while simulations \cite{Peltola2023} and $R_\textrm{int}$-measurements \cite{Gosewich2021_J} indicate its introduction rate being more prominent in a radiation environment involving hadrons.

Thus, the combined observations before and after irradiation indicate that the inter-electrode isolation and consequently position resolution between $n^+$-electrodes without isolation implants can be reached and maintained by the application of sufficient $V_\textrm{bias}$, with the tolerance for high net $N_\textrm{ox}$-levels increasing with reduced active thickness of the sensor. 

%% file: Summary.tex
Measured $CV$-characteristics of 6-inch 300-$\upmu\textrm{m}$-thick $n$-on-$p$ Si-wafer test diodes with floating guard-ring display a double-slope with a distinct $V_\textrm{th,iso}$ in the dynamic region of $C^{-2}V$-curve.
The corresponding TCAD-simulations show that this is due to the electron removal along the Si/SiO$_2$-interface from the inter-electrode gap to the positively biased test-diode pad, until at $V~{\geq}~V_\textrm{th,iso}$ the conduction channel between the two $n^+$-electrodes is broken. This is reflected in the $CV$-characteristics of the test-diode pad as an abrupt drop of capacitance. 
The field-region $N_\textrm{ox}$ is obtained by tuning its simulation input value
until a $V_\textrm{th,iso}$ matching with the measurement is reproduced. 
The values of $N_\textrm{ox}$ determined by this approach agree within uncertainty with $N_\textrm{ox}$ extracted from the measured $V_\textrm{fb}$ of MOS-capacitors from the same samples. $R_\textrm{int}$-simulations show that the threshold voltage where segmented $n^+$-electrodes without isolation implants reach high levels of $R_\textrm{int}$ and thus become electrically isolated is identical to $V_\textrm{th,iso}$($CV$), i.e., $V_\textrm{th,iso}(CV)=V_\textrm{th,iso}(R_\textrm{int})$. Hence, observation of $V_\textrm{th,iso}$($CV$) provides simultaneously information on both field-region $N_\textrm{ox}$ and isolation between $n^+$-electrodes without isolation implants. Further test-diode simulations show that $V_\textrm{th,iso}$ also has significant dependence on both active thickness and bulk doping of the sensor. 

Measured and simulated $CV$-characteristics of the biased guard-ring on the 8-inch HGCAL-prototype and pre-series sensors with varied Si-bulk and passivation oxide parameters 
reproduce the observation of $V_\textrm{th,iso}$ universally. 
These display low values in the range of about $40-100~\textrm{V}$ 
that correspond to field-region $N_\textrm{ox}\approx(5-7)\times10^{10}~\textrm{cm}^{-2}$. 

The apparent high isolation levels between $n^+$-electrodes without isolation implants reachable by the application of sufficient $V_\textrm{bias}$ ($<200~\textrm{V}$ with the initial $N_\textrm{ox}\leq1\times10^{11}~\textrm{cm}^{-2}$) both before and after irradiation in a hadron-dominated radiation environment make this configuration, with a similar 
number of lithography and ion-implantation steps as $p$-on-$n$ sensors, a potentially feasible $n$-on-$p$ sensor candidate for future HEP-experiments. 
With respect to sensor performance, the isolation implantless configuration of position sensitive $n$-on-$p$ sensors removes the probability of discharges or avalanche effects due to excessive electric fields at the $p$-stops.

%% file: appendices.tex
\section{$V_\textrm{fd}$ and bulk doping of the HGCAL sensors}
\label{App1}
\begin{figure*}[htb]
     \centering
     \subfloat[]{\includegraphics[width=.49\textwidth]{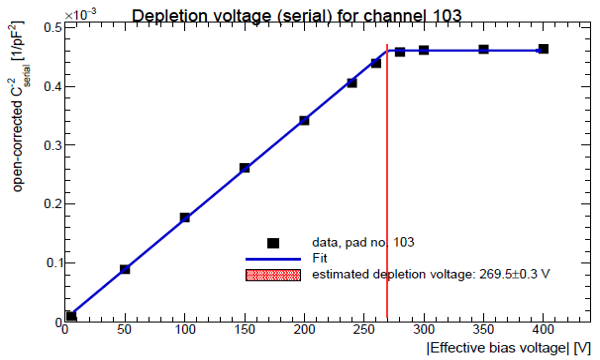}\label{CV_300um_full_oxC}}\hspace{1mm}%
      \subfloat[]{\includegraphics[width=.49\textwidth]{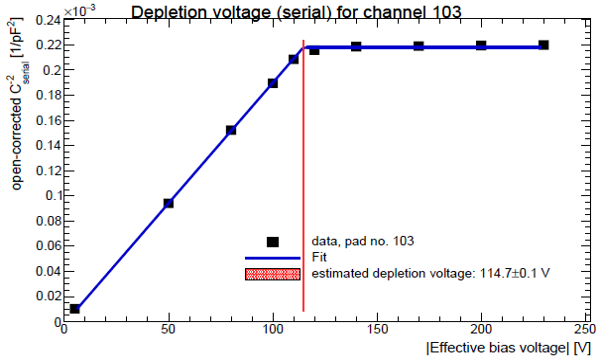}\label{CV_200um_full_oxC}}\\
      \subfloat[]{\includegraphics[width=.49\textwidth]{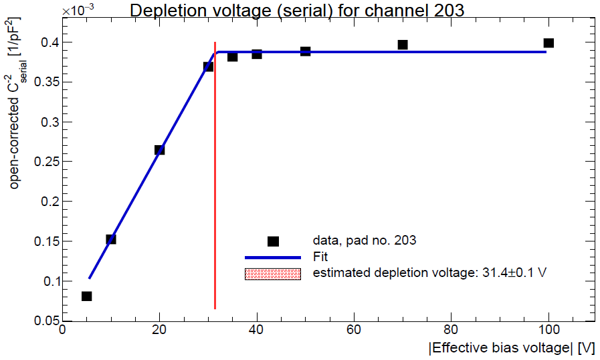}\label{CV_120um_full_oxC}}
    \caption{\small Measured $CV$-characteristics for one representative channel in full 8-inch HGCAL sensors with oxide type `C'. The extracted $V_\textrm{fd}$ is indicated at the crossing-point of the two linear fits to the dynamic- and static-regions of the $C^{-2}V$-data. (a) 300-$\upmu\textrm{m}$-thick LD-sensor with global $\overline{V}_\textrm{fd}=255.7~\textrm{V}\pm1.1\%$ and derived $\overline{N}_\textrm{B}=3.67\times10^{12}~\textrm{cm}^{-3}\pm1.1\%$. (b) 200-$\upmu$m-thick LD-sensor with global $\overline{V}_\textrm{fd}=115.7~\textrm{V}\pm0.6\%$ and derived $\overline{N}_\textrm{B}=3.73\times10^{12}~\textrm{cm}^{-3}\pm0.6\%$. (c) 120-$\upmu$m-thick HD-sensor with global $\overline{V}_\textrm{fd}=31.2~\textrm{V}\pm0.8\%$ and derived $\overline{N}_\textrm{B}=2.80\times10^{12}~\textrm{cm}^{-3}\pm0.8\%$. 
}
\label{Vfd_Nb_HGCALfull}
\end{figure*}

\begin{figure*}[htb]
     \centering
     \subfloat[]{\includegraphics[width=.49\textwidth]{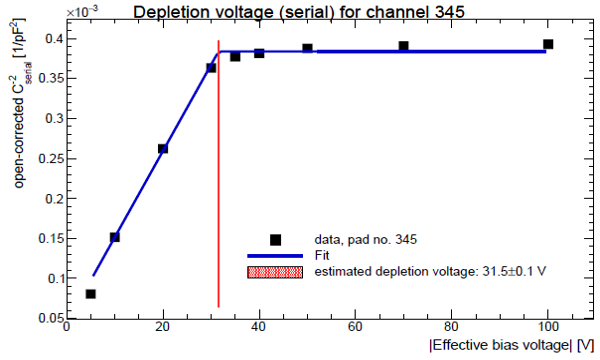}\label{CV_120um_MGW_oxB}}\hspace{1mm}%
     \subfloat[]{\includegraphics[width=.49\textwidth]{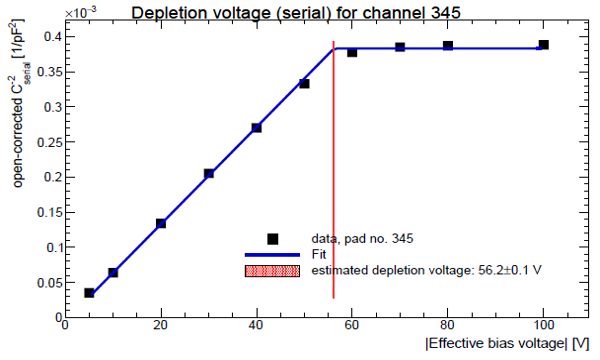}\label{CV_120um_MGW_oxD}}
    \caption{\small Measured $CV$-characteristics for one representative channel in 8-inch partial (`bottom-cut') 120-$\upmu$m HD HGCAL sensors. 
    (a). Sensor with oxide type `B', and with global $\overline{V}_\textrm{fd}=31.2~\textrm{V}\pm0.7\%$ and derived $\overline{N}_\textrm{B}=2.80\times10^{12}~\textrm{cm}^{-3}\pm0.7\%$. (b) Sensor with oxide type `D', and with global $\overline{V}_\textrm{fd}=55.8~\textrm{V}\pm1.0\%$ and derived $\overline{N}_\textrm{B}=5.0\times10^{12}~\textrm{cm}^{-3}\pm1.0\%$.
}
\label{Vfd_Nb_HGCAL_MGW}
\end{figure*}

Data analysis in Figures~\ref{Vfd_Nb_HGCALfull} and~\ref{Vfd_Nb_HGCAL_MGW} was realized by HGCAL Analysis Workflow software\footnote{https://gitlab.cern.ch/CLICdp/HGCAL/}.
%
%
%